\pdfoutput=1
\documentclass[12pt,preprint,url]{aastex}
\usepackage{amsbsy}
\usepackage{amsmath}
\usepackage{lineno}
\newcommand{\be}{\begin{equation}}
\newcommand{\ee}{\end{equation}}

\def\ltsima{$\; \buildrel < \over \sim \;$}

\def\lsim{\lower.5ex\hbox{\ltsima}}
\def\gtsima{$\; \buildrel > \over \sim \;$}
\def\gsim{\lower.5ex\hbox{\gtsima}}

\shorttitle{Magnetospheric fluctuations in accreting pulsars}
\shortauthors{Melatos et al.}

\begin{document}
\title{Tracking hidden magnetospheric fluctuations in accretion-powered pulsars with a Kalman filter}

\author{A. Melatos\altaffilmark{1,2} 
 and N. J. O'Neill\altaffilmark{1,2} 
 and P. M. Meyers\altaffilmark{1,2,3}
 and J. O'Leary\altaffilmark{1,2}}

\email{amelatos@unimelb.edu.au}

\altaffiltext{1}{School of Physics, University of Melbourne,
 Parkville, VIC 3010, Australia}

\altaffiltext{2}{Australian Research Council Centre of Excellence
 for Gravitational Wave Discovery (OzGrav),
 Parkville, VIC 3010, Australia}

\altaffiltext{3}{Theoretical Astrophysics Group,
 California Institute of Technology,
 Pasadena, CA 91125, USA}

\begin{abstract}
\noindent 
X-ray flux and pulse period fluctuations
in an accretion-powered pulsar convey important information 
about the disk-magnetosphere interaction.
It is shown that simultaneous flux and period measurements
can be analysed with a Kalman filter 
based on the standard magnetocentrifugal accretion torque 
to generate accurate time-dependent estimates
of three hidden state variables,
which fluctuate stochastically and cannot be measured directly:
the mass accretion rate,
the Maxwell stress at the disk-magnetosphere boundary,
and the radiative efficiency of accretion onto the stellar surface.
The inferred fluctuation statistics carry implications for the physics of
hydromagnetic instabilities at the disk-magnetosphere boundary
and searches for continuous gravitational radiation from 
low-mass X-ray binaries.
\end{abstract}


\section{Introduction 
 \label{sec:kal1}}
Coherent X-ray timing of accretion-powered pulsars
affords insights into how the angular velocity $\Omega(t)$ of the
neutron star and its time derivative $\dot{\Omega}(t)$
fluctuate stochastically in response to the hydromagnetic accretion torque.
Timing data from satellite missions such as the {\em Compton Gamma Ray Observatory},
{\em Rossi X-Ray Timing Explorer (RXTE)},
and {\em Neutron Star Interior Composition Explorer (NICER)}
\citep{gen16}
reveal a number of intriguing, torque-related phenomena,
including torque reversals, 
pulse profile phase delays at the fundamental frequency and its harmonics,
phase-flux correlations,
and red noise in the power spectral density (PSD) of $\Omega(t)$ and $\dot{\Omega}(t)$
\citep{bil97,rig08,yan17,pat21,ser21}.
Simultaneously, 
high-time-resolution measurements of fluctuations in the aperiodic X-ray luminosity $L(t)$
shed light on the physics of the accretion disk and disk-magnetosphere boundary.
Observed phenomena include nonstationarity,
red noise in the PSD of $L(t)$ (typically a doubly broken power-law),
quasiperiodic oscillations whose frequencies depend on flux,
and a linear scaling between flux and root-mean-square variability,
which extends to other accreting compact objects such as active galaxies
\citep{utt01,rev15,pat21,dem22}.

In principle, $\Omega(t)$ and $L(t)$ fluctuations can be related theoretically,
as both variables depend on the mass accretion rate
under the canonical magnetocentrifugal hypothesis
\citep{gho79}.
Magnetocentrifugal accretion is consistent with phenomenological properties
of accretion-powered pulsars at the population level and in individual objects,
e.g.\ the recycling scneario confirmed by the discovery of SAX J1808.4$-$3658,
and the spin-up line for radio millisecond pulsars in the $\Omega$-$\dot{\Omega}$ plane
\citep{pat21}.
However the disk-magnetosphere interaction is complicated geometrically and hydromagnetically;
three-dimensional simulations predict the emergence of twisted magnetic field structures 
disrupted episodically by instabilities at the disk-magnetosphere boundary
\citep{rom03,rom05,kul08}.
Consequently it is challenging to reconcile the measured time series 
$\Omega(t)$ and $L(t)$ in detail in individual objects,
predict correlations like $\langle \Omega(t) L(t') \rangle$,
and predict the origin and onset of phenomena such as quasiperiodic oscillations.
Progress has occurred in certain directions,
e.g.\
employing autoregressive moving average models to study torque-luminosity correlations
of the form $\langle \dot{\Omega}(t) L(t') \rangle$
\citep{bay93},
and testing for consistency with random walk and shot noise processes
\citep{dek93,bay97,laz97}.
Evidence has also been reported for a break in the power-law PSD of $L(t)$
near the neutron star spin frequency (and hence the Kepler frequency
at the disk-magnetosphere boundary) for accretion-powered X-ray pulsars
near magnetocentrifugal equilibrium
\citep{rev09,rev15,mon22}.
Torque-luminosity modeling has been performed on GRO J1744$-$28 and 2S 1417$-$624,
using a phemomenological torque-luminosity scaling of power-law form
\citep{san17,ser21}.

Understanding the time-dependent connection between $\Omega(t)$ and $L(t)$
has become an imperative recently in another field:
gravitational wave astronomy.
Searches for continuous, quasimonochromatic, gravitational wave signals 
from accretion-powered pulsars are a priority for long-baseline detectors such as the
Laser Interferometer Gravitational Wave Observatory (LIGO)
\citep{mid20,abb21}.
The target list extends beyond accretion-powered pulsars to embrace
nonpulsating accreting neutron stars in low-mass X-ray binaries,
including sources that exhibit thermonuclear burst oscillations 
and kilohertz quasiperiodic oscillations,
where $\Omega(t)$ is measured intermittently or not at all
\citep{wat08,ril13,ril22}.
The PSD of $\Omega(t)$ affects the gravitational wave sensitivity, 
because it limits the coherence time $T_{\rm drift}$ of a search with an optimal matched filter,
which assumes that $\Omega(t)$ evolves deterministically
\citep{wat08,ril13,muk18,ril22}.
Semicoherent algorithms have been developed to track stochastic fluctuations in $\Omega(t)$
\citep{goe11,whe15,suv17,mel21}
but they require $T_{\rm drift}$ to be known in advance.
In nonpulsating systems, where $\Omega(t)$ cannot be measured directly,
it would be advantageous to infer $T_{\rm drift}$
from luminosity fluctuations, i.e.\ from the observed PSD of $L(t)$.
A pioneering study with this goal was completed by \citet{muk18},
but work remains to be done.
A general theoretical understanding of the connection between $\Omega(t)$ and $L(t)$
would prove valuable in this context.

In this paper, we demonstrate a new, self-consistent, signal processing framework 
based on a Kalman filter
to relate simultaneous observations of the time series $\Omega(t)$ and $L(t)$
to the canonical dynamical model of magnetocentrifugal accretion
and track the evolution of hidden state variables of physical interest,
such as the Maxwell stress at the disk-magnetosphere boundary.
The Kalman filter framework extends previous studies in three ways.
(i) It connects $\Omega(t)$ and $L(t)$ through a specific physical model of accretion,
namely the magnetocentrifugal model, instead of invoking a generic random process
(e.g.\ shot noise) or a phenomenological torque-luminosity scaling (e.g.\ power law).
(ii) It tracks the progress of the system through the most likely sequence
of hidden states consistent with the specific time-ordered $\Omega(t)$ and $L(t)$ data
observed, instead of ensemble-averaged statistics such as the associated PSDs.
(iii) It infers the statistics of $\Omega(t)$, $L(t)$, and the hidden state variables
simultaneously, rather than constructing PSDs for $\Omega(t)$ and $L(t)$
before relating them to the underlying accretion dynamics.
The paper shares some features with recent work to estimate,
using a Kalman filter, 
the parameters of the classic, two-component, crust-superfluid model 
of a neutron star driven by timing noise
\citep{bay91,mey21b},
although the latter reference does not analyse $L(t)$;
see also \citet{mey21a}.
The paper also shares some features with recent work applying
continuous-time autoregressive moving average (CARMA) models and their variants
to a diverse selection of astrophysical massive time-domain data sets,
including accretion-powered pulsars, active galaxies, and variable stars
\citep{kel14,hu20,elo21}.
Indeed, by way of illustration,
\citet{kel14} analysed {\em RXTE} data from the low-mass X-ray binary XTE 1550$-$564,
whose compact object is a black hole rather than a neutron star,
and concluded that its $L(t)$ fluctuations are consistent with a 
CARMA process of order $(5,4)$.
The analysis does not consider $\Omega(t)$,
because $\Omega(t)$ cannot be measured for a black hole,
and the generic ${\rm CARMA}(5,4)$ process is not derived from a
physical model of accretion.

The paper is structured as follows.
In \S\ref{sec:kal2} we introduce the stochastic differential equations of motion,
which govern how the state variables describing magnetocentrifugal accretion evolve,
as well as the measurement equations,
which relate the observables $\Omega(t)$ and $L(t)$ to the state variables,
some of which are hidden.
The equations are linearized about magnetocentrifugal equilibrium
to prepare for implementing a Kalman filter.
In \S\ref{sec:kal3} we set out a practical recipe for estimating
the evolution of the state variables using a Kalman filter given the observed time series
$\Omega(t)$ and $L(t)$.
Validation tests with synthetic data are presented in \S\ref{sec:kal4},
and the accuracy of parameter estimation is quantified approximately.
Astrophysical implications are canvassed briefly in \S\ref{sec:kal5}.
Preliminary extensions of the analysis to generalized models
of magnetocentrifugal accretion are introduced in the appendices.
Applying the method to real astronomical data is postponed to future work,
to be undertaken in collaboration with the X-ray timing community,
which enjoys access to high-quality, calibrated data
and specialized analysis software.

\section{Accretion dynamics
 \label{sec:kal2}}
Accretion from a disk onto a magnetized neutron star is a complicated,
time-dependent processs involving nonlinear feedback between the disk and
magnetosphere,
mediated by hydromagnetic instabilities at the disk-magnetosphere boundary,
as seen in three-dimensional numerical simulations
\citep{rom03,rom05,kul08}.
The time series $\Omega(t)$ and $L(t)$ supplied by X-ray timing experiments
do not contain enough information to infer uniquely the spatial structure 
in the simulations,
e.g.\ the magnetic field geometry near the disk-magnetosphere boundary.
In this paper, therefore, we model accretion in a spatially averaged manner 
within the successful magnetocentrifugal paradigm
\citep{gho79}.
In \S\ref{sec:kal2a},
we define and relate the two observables and four state variables
that constitute the model,
three of which are hidden,
including the Maxwell stress at the disk-magnetosphere boundary,
which is important physically and notoriously difficult to measure
\citep{pat21}.
The canonical magnetocentrifugal torque is written down in \S\ref{sec:kal2b}.
The state of rotational equilibrium, about which the system fluctuates,
and the stochastic driving forces which drive the system away from equilibrium
are specified in \S\ref{sec:kal2c} and \S\ref{sec:kal2d} respectively.
Linearized versions of the dynamical equations for the state variables
and the measurement equations relating the observables to the state variables
are presented in \S\ref{sec:kal2e}.
The linearized equations are suitable for analysis with a Kalman filter.

\subsection{Observables and state variables
 \label{sec:kal2a}}
Consider a hypothetical X-ray timing experiment targeting an accretion-powered pulsar.
The experiment returns raw photon times of arrival,
which are barycentered and converted into time series of the pulse period,
$P(t_1),\dots,P(t_N)$,
and the aperiodic X-ray luminosity,
$L(t_1),\dots,L(t_N)$,
using standard coherent timing methods, e.g.\
Fourier decomposition of the pulse profile and pulse folding
\citep{bil97,pat21,ser21}.
\footnote{
The luminosity $L(t)$ is discussed for notational convenience in this paper.
In practice, the analysis does not rely on knowing the distance $D$ to the source.
One can work instead with the aperiodic X-ray flux $F_X(t)= L(t)/(4\pi D^2)$
and rescale the parameters to be estimated accordingly by powers of $D$.
\label{foot:kal1}
}
The time series are sampled simultaneously at $N$ epochs 
$t_1 \leq  \dots \leq t_N$,
spaced regularly or irregularly,
during the interval $0\leq t \leq T_{\rm obs}$.
\footnote{
The signal processing framework in this paper can be generalized to handle
nonsimultaneous sampling of the two time series,
if there is demand in the future
\citep{gel74}.
}

Consider also the standard magnetocentrifugal model of disk accretion
\citep{gho79}.
Reformulated slightly from its canonical form,
the model can be written in terms of four state variables,
which are functions of time $t$.
One state variable, the angular velocity $\Omega(t)$ of the neutron star,
is related closely to $P(t)$, as discussed below.
The other three state variables are hidden;
they are related indirectly to $P(t)$ and $L(t)$ and cannot be measured directly.
Let $S(t)$ denote the Maxwell stress at the disk-magnetsophere boundary,
which opposes the ram pressure of the radially inflowing disk material
(units: ${\rm g\,cm^{-1} \, s^{-2}}$).
Let $Q(t)$ be the rate at which mass flows from the accretion disk
into the disk-magnetosphere boundary (units: ${\rm g\,s^{-1}}$).
Let $\eta(t)$ be the efficiency with which the gravitational potential energy
of material falling onto the stellar surface is converted into X-rays
(units: dimensionless).
The roles played by $S(t)$, $Q(t)$, and $\eta(t)$ in the accretion dynamics
are defined in \S\ref{sec:kal2b} and \S\ref{sec:kal2d}.
All quantities are expressed in CGS units.

The first step in formulating a Kalman filter is to relate the observables
and state variables.
The angular velocity of the star is essentially measured directly, viz.
\begin{equation}
 P(t) = 2\pi / \Omega(t) + N_P(t)~.
\label{eq:kal1}
\end{equation}
The additive measurement noise $N_P(t)$
is assumed to be Gaussian and white,
with
$\langle N_P(t_n) \rangle = 0$
and
$\langle N_P(t_n) N_P(t_{n'}) \rangle = \Sigma_{PP}^2 \delta_{n,n'}$,
where $\delta_{n,n'}$ denotes the Kronecker delta.
Note that $\Sigma_{PP}$ has units of ${\rm s}$.
Equation (\ref{eq:kal1}) is a reasonable starting point for estimating
the parameters of the magnetocentrifugal model.
However it is an approximation;
the discovery of (anti)correlations between pulse phase residuals and X-ray flux 
(instead of $\dot{\Omega}$ and X-ray flux) in several accretion-powered pulsars
\citep{pat09}
implies that the pulse frequency derivative does not equal
the spin frequency derivative exactly
\citep{rig08,pat21}.
That is, one has
$\langle \Omega(t) \rangle \neq 2\pi/\langle P(t) \rangle$ in some circumstances,
contradicting (\ref{eq:kal1}),
perhaps due to secular hot spot migration driven by disk-magnetosphere instabilities
\citep{rom03,kul08}.

The measured X-ray luminosity can be related indirectly to the mass accretion rate
and radiative efficiency by the usual energy conservation formula, viz.\
\begin{equation}
 L(t) = G M Q(t) \eta(t) / R + N_L(t)~.
\label{eq:kal2}
\end{equation}
The additive measurement noise $N_L(t)$ is assumed to be Gaussian and white,
with
$\langle N_L(t_n) \rangle = 0$
and
$\langle N_L(t_n) N_L(t_{n'}) \rangle = \Sigma_{LL}^2 \delta_{n,n'}$.
If there is reason to believe that $N_P(t_n)$ and $N_L(t_n)$
are correlated through the measurement process,
it is straightforward to generalize the analysis
to accommodate $\langle N_P(t_n) N_L(t_{n'}) \rangle = \Sigma_{PL}^2 \delta_{n,n'} \neq 0$.
In (\ref{eq:kal2}),
$\eta(t)$ is the fraction of the specific gravitational potential energy $GM/R$
converted to X-rays, when infalling matter strikes the stellar surface,
$M$ and $R$ denote the mass and radius of the star respectively,
and $G$ denotes Newton's gravitational constant.
A one-to-one correspondence is assumed sometimes between $L(t)$ and $Q(t)$,
with $\eta(t)=1$ in (\ref{eq:kal2}).
In this paper we allow $\eta(t)$ to fluctuate in the range $0 < \eta(t) < 1$
\citep{san17}.
Partly the fluctuations arise from radiative processes:
how much of the gravitational potential energy is converted into heat
and hence X-rays?
Partly they arise from nonconservative mass transfer:
how much of $Q(t)$ lands on the stellar surface, and how much is directed into an outflow
\citep{mar19}?
An idealized, phenomenological model 
of the $\eta(t)$ dynamics is introduced in \S\ref{sec:kal2d}.

\subsection{Magnetocentrifugal torque
 \label{sec:kal2b}}
In the canonical magnetocentrifugal picture of accretion \citep{gho79},
there are two characteristic radii.
The Alfv\'{en} radius, $R_{\rm m}(t)$,
defines the disk-magnetosphere boundary.
It is located where the magnetospheric Maxwell stress balances the disk ram pressure,
viz.\ $S \approx \rho v^2$,
where $\rho = Q/(4\pi R_{\rm m}^2 v)$
and $v = (GM/R_{\rm m})^{1/2}$
are the mass density and infall speed respectively
in a cylindrically symmetric inflow approximately in free fall.
In terms of the hidden state variables,
the stress balance condition translates to
\begin{equation}
 R_{\rm m}(t)
 =
 (4\pi)^{-2/5} (GM)^{1/5} Q(t)^{2/5} S(t)^{-2/5}~.
\label{eq:kal3}
\end{equation}
The corotation radius, $R_{\rm c}(t)$, is located where the Kepler frequency equals
the angular velocity of the star, viz.\
\begin{equation}
 R_{\rm c}(t)
 =
 (GM)^{1/3} \Omega(t)^{-2/3}~.
\label{eq:kal4}
\end{equation}

The fastness parameter $(R_{\rm m}/R_{\rm c})^{3/2}$ controls the sign
of the magnetocentrifugal torque on the star.
For $R_{\rm m} < R_{\rm c}$, material at the disk-magnetosphere boundary
orbits faster than the star rotates.
Some fraction [related but not equal to $\eta(t)$] falls onto the star 
and spins it up through a combination of hydromagnetic and mechanical torques.
For $R_{\rm m} > R_{\rm c}$, in the propeller phase, 
the material at the disk-magnetosphere boundary orbits slower than the star rotates.
It is flung outwards centrifugally by the corotating magnetosphere
and spins down the star.
Importantly, some material can still accrete onto the stellar surface
during the propeller phase for $1 \lesssim R_{\rm m}/R_{\rm c} \lesssim 3$,
even if the remainder is diverted into an outflow
\citep{pap15,mar19,pat21}.
Consequently the net torque involves both hydromagnetic
and mechanical components in general.
The equation of motion for $\Omega(t)$ then reads
\begin{equation}
 \frac{d\Omega}{dt}
  =
 \frac{(GM)^{1/2}}{I}
 \left\{ 1 - \left[ \frac{R_{\rm m}(t)}{R_{\rm c}(t)} \right]^{3/2} \right\} 
 R_{\rm m}(t)^{1/2} Q(t)~,
\label{eq:kal5}
\end{equation}
where $I$ denotes the star's moment of inertia.

In this paper, for the sake of simplicity,
we neglect modifications of (\ref{eq:kal5}) due to radiation pressure
\citep{and05,has15b},
magnetic dipole braking in quiescence
\citep{pat10,pap11,mel16},
and gravitational radiation reaction
\citep{bil98,mel05}.
The modifications are straightforward to make,
whenever the data are detailed enough to warrant their inclusion.
We also neglect disk warping and precession,
caused by misalignment between the magnetic and rotation axes of the star
and the angular momentum vector of the disk
\citep{fou11,lai14,rom21};
see Figures 6 and 11 in \citet{rom21} for a vivid illustration of how
warping and precession cause variability in $Q(t)$.
Misalignment necessarily leads to complicated, three-dimensional flows
\citep{rom21},
whose description lies outside the scope of this paper.

\subsection{Magnetocentrifugal equilibrium
 \label{sec:kal2c}}
Rotational equilibrium corresponds to zero torque,
which is achieved for $R_{\rm c}(t)=R_{\rm m}(t) = R_{{\rm m}0} = {\rm constant}$,
$S(t)=S_0={\rm constant}$,
$Q(t)=Q_0={\rm constant}$,
$\eta(t)=\eta_0={\rm constant}$,
and
$\Omega(t)=\Omega_0={\rm constant}$, with
\begin{equation}
 \Omega_0
 = 
 (4\pi)^{3/5} (GM)^{1/5} Q_0^{-3/5} S_0^{3/5}
\label{eq:kal6}
\end{equation}
and
\begin{equation}
 L_0
 =
 GM Q_0 \eta_0 / R~.
\label{eq:kal7}
\end{equation}

Often it is useful to write $S_0$ and $R_{{\rm m}0}$
in terms of the star's magnetic moment $\mu$
(units: ${\rm G \, cm^3}$),
assuming a dipole magnetic field and hence
$S_0 = (2\pi)^{-1} \mu^2 R_{{\rm m}0}^{-6}$
inside the magnetosphere. The results are
$S_0 = 2^{-19/7} \pi^{-1} (GM)^{6/7} \mu^{-10/7} Q_0^{12/7}$,
$R_{{\rm m}0} = 2^{2/7} (GM)^{-1/7} \mu^{4/7} Q_0^{-2/7}$
(a familiar expression in the literature),
and hence
\begin{equation}
 \mu
 =
 2^{-1/2} (GM)^{5/6} \Omega_0^{-7/6} Q_0^{1/2}~.
\label{eq:kal8}
\end{equation}
In this paper, we assume that $\mu$ is constant for simplicity. 
However there is theoretical \citep{shi89,pay04,zha06}
and observational \citep{pat12} evidence, that polar magnetic burial
reduces $\mu$ in the short term, during accretion episodes,
and in the long term,
as reflected in the spin distribution of accretion-powered pulsars
\citep{pri11,wan11}.

The four-dimensional state vector in equilibrium, $(\Omega_0, Q_0, S_0, \eta_0)$,
is a key input into the Kalman filter, as described in \S\ref{sec:kal3}.
However, the two time-averaged observables 
$\Omega_0 = \langle 2\pi/ P(t) \rangle$ 
and 
$L_0=\langle L(t) \rangle$,
and the magnetocentrifugal equilibrium condition $R_{\rm c}(t)=R_{\rm m}(t)$,
contain only three independent pieces of information.
One is therefore left with two approaches to solve for 
the four components of $(\Omega_0, Q_0, S_0, \eta_0)$:
(I) assume a plausible theoretical value for one component,
e.g.\ $\eta_0=0.5$;
or (II) exploit the time-dependent information in $P(t)$ and $L(t)$,
not just $\langle P(t) \rangle$ and $\langle L(t) \rangle$,
and include one component in the Kalman filter
as an unknown to be estimated.
Approach II is preferable as it is more general.
Handy formulas for the components of $(\Omega_0, Q_0, S_0, \eta_0)$
following either approach
are presented in Appendix \ref{sec:kalappb} for the convenience of the reader.

Many accretion-powered pulsars exist in a state of disequilibrium
\citep{bil97,yan17,mus22,ser22}.
They spin up or down secularly over long intervals, typically lasting years,
with
$| R_{\rm c}(t) - R_{\rm m}(t) | \gtrsim R_{\rm c}(t)$.
Examples include Her X$-$1 and 4U 1626$-$67, 
which accrete via Roche lobe overflow,
where the secular intervals last $\gtrsim 10\,{\rm yr}$
\citep{ser22}.
For such systems, it is inappropriate to linearize (\ref{eq:kal5}) 
about $R_{\rm c}(t) = R_{\rm m}(t) = R_{\rm m0}$,
as in \S\ref{sec:kal2e},
in order to apply a linear Kalman filter.
Instead, in the absence of an accepted analytic theory of torque transitions
\citep{nel97,van98,loc04,lai14,gen22},
it makes sense to analyze each decade-long secular interval separately,
by applying a nonlinear (e.g.\ unscented) Kalman filter
\citep{jul97}
to (\ref{eq:kal5}) in its nonlinear form.
An illustrative worked example is presented in Appendix \ref{sec:kalappc}
for completeness.

\subsection{Stochastic fluctuations
 \label{sec:kal2d}}
Accretion-powered pulsars are stochastic systems.
Partly the stochasticity is driven externally,
e.g.\ flicker noise due to propagating fluctuations 
in the disk $\alpha$ parameter on the viscous time-scale
\citep{lyu97},
or longer-term $Q(t)$ modulation as the companion star evolves.
Partly the stochasticity emerges internally due to nonlinear feedback loops,
e.g.\ self-healing Rayleigh-Taylor instabilities at the disk-magnetosphere boundary
\citep{rom03,kul08,das22},
or cyclic accretion due to disk trapping at the magnetocentrifugal barrier
\citep{dan12}.
In this paper, we focus on internal stochasticity,
specifically mean-reverting fluctuations around magnetocentrifugal equilibrium.

Hydromagnetic processes at the disk-magnetosphere boundary cannot be observed directly.
The physics depends on the spatial structure of the magnetic field, 
which is too complicated to be inferred uniquely from
X-ray timing measurements of $P(t)$ and $L(t)$.
Three-dimensional hydromagnetic simulations are expensive computationally
and cannot be repeated often enough to predict ensemble statistics
of hidden state variables reliably, e.g.\ the PSD of $S(t)$
\citep{rom03,kul08}.
Accordingly, we adopt an idealized, phenomenological dynamical model,
in which the hidden state variables $Q(t)$, $S(t)$, and $\eta(t)$
execute mean-reverting random walks driven by white noise
\citep{dek93}.
That is, $Q(t)$, $S(t)$, and $\eta(t)$ satisfy the Langevin equations
\begin{eqnarray}
 \frac{dQ}{dt}
 & = & 
 -\gamma_Q [Q(t) -Q_0] + \xi_Q(t)~,
\label{eq:kal9}
 \\
 \frac{dS}{dt}
 & = & 
 -\gamma_S [S(t) -S_0]  + \xi_S(t)~,
\label{eq:kal10}
 \\
 \frac{d\eta}{dt}
 & = & 
 -\gamma_\eta [\eta(t) -\eta_0] +  \xi_\eta(t)~,
\label{eq:kal11}
\end{eqnarray}
where $\gamma_Q^{-1}$, $\gamma_S^{-1}$, and $\gamma_\eta^{-1}$
are characteristic time-scales of mean reversion,
and $\xi_Q(t)$, $\xi_S(t)$, and $\xi_\eta(t)$
are white-noise driving terms with ensemble statistics
$\langle \xi_{A}(t) \rangle = 0$
and
\begin{equation}
 \langle \xi_{A}(t) \xi_{A'}(t') \rangle 
 =
 \sigma_{AA'}^2 \delta(t-t')~,
\label{eq:kal12}
\end{equation}
i.e.\ delta-correlated in time, with $A, A' \in \{ Q,S,\eta \}$.
\footnote{
The equations of motion (\ref{eq:kal9})--(\ref{eq:kal11}) 
describe a continuous-time, Ornstein-Uhlenbeck process
\citep{gar94},
whereas the measurement equations (\ref{eq:kal1}) and (\ref{eq:kal2})
are sampled at the discrete epochs $t_1 \leq \dots \leq t_N$.
Hence $\langle \xi_A(t) \xi_{A'}(t') \rangle$ in (\ref{eq:kal12})
is proportional to the Dirac delta function $\delta(t-t')$,
whereas $\langle N_B(t_n) N_{B'}(t_{n'}) \rangle$
in \S\ref{sec:kal2a} is proportional to the Kronecker delta $\delta_{n,n'}$
for $B, B' \in \{P, L \}$.
The units of $\sigma_{AA'}^2$ in (\ref{eq:kal12})
are the units of $A$ multiplied by the units of $A'$ divided by seconds,
whereas the units of $\Sigma_{BB'}^2$ in \S\ref{sec:kal2a}
are the units of $B$ multiplied by the units of $B'$.
}
In this paper, we assume $\sigma_{AA'}=0$ for $A\neq A'$
for the purpose of illustration,
and to keep the number of unknown parameters manageable.
Cross-correlations of the form
$\sigma_{A A'} \neq 0$ for $A\neq A'$
are straightforward to add in the future, 
if astronomical data demand their inclusion.

The Langevin equations (\ref{eq:kal9})--(\ref{eq:kal12}) ensure that $Q(t)$, $S(t)$, and $\eta(t)$
wander randomly about their equilibrium values without drifting secularly,
with root-mean-square fluctuations $\sim \gamma_{Q}^{-1/2} \sigma_{QQ}$,
$\gamma_{S}^{-1/2} \sigma_{SS}$, and $\gamma_{\eta}^{-1/2} \sigma_{\eta\eta}$ respectively.
For example,
the probability density function for $Q(t)$ in the limit $t\rightarrow \infty$
takes the form $p(Q) \propto \exp[-\gamma_Q (Q-Q_0)^2/\sigma_{QQ}^2]$.
Analogous formulas apply for $p(S)$ and $p(\eta)$.
Formally (\ref{eq:kal9})--(\ref{eq:kal12}) allow for 
unphysical fluctuations with $Q$, $S$, $\eta < 0$ or $\eta > 1$.
In practice, however, the probabilities of such fluctuations are exponentially small,
as one has $\gamma_Q^{-1/2} \sigma_{QQ} \ll Q_0$,
$\gamma_S^{-1/2} \sigma_{SS} \ll S_0$, and $\gamma_\eta^{-1/2} \sigma_{\eta\eta} \ll \eta_0$
in accretion-powered pulsars in their active phase.
There is no observational evidence that the disk-magnetosphere system is disrupted catastrophically,
e.g.\ $Q<0$ (accretion ceases) or $S<0$ (Maxwell stress vanishes).
Systems in quiescence are not considered here
\citep{pat21}.
The approximation (\ref{eq:kal9})--(\ref{eq:kal12})
simplifies the Kalman filter considerably (see \S\ref{sec:kal3}).

We emphasize that (\ref{eq:kal9})--(\ref{eq:kal12}) are highly idealized
in important respects.
For example, it is likely that $\xi_S(t)$ and $\xi_\eta(t)$ are anticorrelated
to some degree, with $\sigma_{S\eta} \neq 0$.
Simulations show that Rayleigh-Taylor instabilities at the disk-magnetosphere boundary
open up transient magnetic channels,
accompanied by fluctuations in $S(t)$,
which temporarily permit ``fingers'' of disk material to break through the magnetosphere 
and strike the stellar surface, 
before the channels close, and equilibrium is restored
\citep{rom03,rom05,kul08}.
This gating process is correlated with the nonradiative component of $\eta(t)$,
i.e.\ the component associated with nonconservative mass transfer 
as discussed in \S\ref{sec:kal2a} and \S\ref{sec:kal2b}.
When the Rayleigh-Taylor channels close,
and $S(t)$ fluctuates above $S_0$,
some fraction of $Q(t)$ is redirected magnetocentrifugally into an outflow,
and $\eta(t)$ fluctuates below $\eta_0$
\citep{mar19,pat21}.
\footnote{
The anticorrelation is imperfect for two reasons.
First, the magnetic geometry at the disk-magnetosphere boundary is complicated,
and $S(t)$ does not always decrease, when a Rayleigh-Taylor channel opens.
Second, $\eta(t)$ contains a radiative component
(the fraction of the gravitational potential energy converted into heat
and hence X-rays, when matter strikes the stellar surface),
which does not depend on $S(t)$.
}
As noted above, it is straightforward to implement $\sigma_{S\eta} \neq 0$
through (\ref{eq:kal12}) in future work,
at the cost of introducing an additional parameter,
if the data warrant.

A second idealization 
is that $\xi_Q(t)$, $\xi_S(t)$, and $\xi_\eta(t)$ obey white noise statistics.
X-ray timing experiments indicate that some objects exhibit
red noise in the torque $\dot{P}(t)$ \citep{bil97,ser22}
and light curve $L(t)$ \citep{muk18}.
We emphasize that red noise is consistent with (\ref{eq:kal9})--(\ref{eq:kal12}):
although $\xi_Q(t)$, $\xi_S(t)$, and $\xi_\eta(t)$ are white,
the PSDs of $Q(t)$, $S(t)$, and $\eta(t)$ are red,
because $\xi_Q(t)$, $\xi_S(t)$, and $\xi_\eta(t)$ 
appear in the derivatives $dQ/dt$, $dS/dt$, and $d\eta/dt$,
and the deterministic terms on the right-hand sides of
(\ref{eq:kal9})--(\ref{eq:kal12}) act as low-pass filters
on the long time-scales 
$\gamma_Q^{-1}$, $\gamma_S^{-1}$, and $\gamma_{\eta}^{-1}$ respectively,
cf.\ \citet{bay93}.
The red noise feeds into $\Omega(t)$ and hence the observables $P(t)$ and $L(t)$
through (\ref{eq:kal1}), (\ref{eq:kal2}), and (\ref{eq:kal5}).
Nonetheless, the shapes (e.g.\ power-law index) of the PSDs of $P(t)$ and $L(t)$ 
are not reproduced in detail by (\ref{eq:kal9})--(\ref{eq:kal12}) in some objects
\citep{ser22}.
We persevere with (\ref{eq:kal9})--(\ref{eq:kal12}) in this introductory paper
by way of illustration,
while noting that the model can be generalized easily, 
when the data demand,
by augmenting (\ref{eq:kal9})--(\ref{eq:kal12}) with additional filters
to generate noise with the desired color
\citep{kel14,hu20,elo21}.

Data volumes available at present ($N \lesssim 10^3$)
are insufficient to constrain dynamical models of the disk-magnetosphere boundary
that are more realistic than (\ref{eq:kal5}) and (\ref{eq:kal9})--(\ref{eq:kal11}).
Yet there is no doubt that (\ref{eq:kal5}) and (\ref{eq:kal9})--(\ref{eq:kal11})
oversimplify many important aspects of the accretion physics,
beyond those highlighted in the previous two paragraphs.
For example, the Maxwell stress is a tensor not a scalar,
forces perpendicular to the disk cannot be neglected,
and the transition from spin up ($R_{\rm m} < R_{\rm c}$)
to spin down ($R_{\rm m} > R_{\rm c}$) occurs more abruptly than implied
by (\ref{eq:kal5}).
We discuss the implications of these and other approximations
in Appendix \ref{sec:kalappd} and sketch out,
for the sake of completeness,
how the Kalman filter framework can be refined to accommodate 
some of these effects in the future, when more data become available.
Specifically, we present generalized versions of
(\ref{eq:kal5}) and (\ref{eq:kal9})--(\ref{eq:kal11}),
that describe an abrupt propeller transition and disk trapping
\citep{dan12,dan17},
albeit still in an idealized form,
and sketch out how to analyze the generalized models with an unscented Kalman filter
\citep{jul97}.

\subsection{Linearized equations of motion and measurement equations
 \label{sec:kal2e}}
The analysis in this paper applies to accretion-powered pulsars
whose fluctuations about magnetocentrifugal equilibrium are small,
with $\gamma_Q^{-1/2} \sigma_{QQ} \ll Q_0$, $\gamma_S^{-1/2} \sigma_{SS} \ll S_0$,
and
$\gamma_\eta^{-1/2} \sigma_{\eta\eta} \ll \eta_0$
as discussed in \S\ref{sec:kal2d}.
Small fluctuations can be treated with a linear Kalman filter
(see \S\ref{sec:kal3}).
Ultimately detailed comparisons with observational data will be needed
to assess whether or not the linear approximation is accurate in individual objects.
Evidence exists for nonlinearity in dynamical models of other accreting systems,
such as black hole binaries \citep{tim00,man16}
and cataclysmic variables
\citep{sca14}.
Quiescent systems,
and catastrophic events such as disk disruption or magnetospheric collapse,
are not considered here.

Let us denote perturbed fractional quantities with the subscript `1'
and unperturbed absolute quantities with the subscript `0' (see \S\ref{sec:kal2c}).
For example, we write $\Omega_1(t) = [\Omega(t)-\Omega_0] / \Omega_0$,
$Q_1(t) = [Q(t)-Q_0] / Q_0$, and so on.
Linearizing the equations of motion
(\ref{eq:kal3})--(\ref{eq:kal5}) and (\ref{eq:kal9})--(\ref{eq:kal11})
yields
\begin{equation}
 \frac{d}{dt}
 \left(
  \begin{tabular}{c}
   $\Omega_1$ \\ $Q_1$ \\ $S_1$ \\ $\eta_1$
  \end{tabular}
 \right)
 =
 \left(
  \begin{tabular}{cccc}
   $-\gamma_\Omega$ & $-3\gamma_\Omega / 5$ & 
    $3\gamma_\Omega / 5$ & 0 \\ 
   $0$ & $ -\gamma_Q$ & $0$ & $0$  \\
   $0$ & $0$ & $-\gamma_S$ & $0$ \\
   $0$ & $0$ & $0$ & $-\gamma_\eta$
  \end{tabular}
 \right)
 \left(
  \begin{tabular}{c}
   $\Omega_1$ \\ $Q_1$ \\ $S_1$ \\ $\eta_1$
  \end{tabular}
 \right)
 +
 \left(
  \begin{tabular}{c}
   $0$ \\ $Q_0^{-1} \xi_Q$ \\ $S_0^{-1} \xi_S$ \\ $\eta_0^{-1} \xi_\eta$
  \end{tabular}
 \right)~,
\label{eq:kal13}
\end{equation}
with
\begin{equation}
 \gamma_\Omega 
 =
 \frac{ (GM)^{1/2} R_{{\rm m}0}^{1/2} Q_0 }{I \Omega_0}~.
\label{eq:kal14}
\end{equation}
Linearizing the measurement equations (\ref{eq:kal1}) and (\ref{eq:kal2}) yields
\begin{equation}
 P_1 = - \Omega_1 + P_0^{-1} N_P
\label{eq:kal15}
\end{equation}
and
\begin{equation}
 L_1 =
 Q_1 + \eta_1 
 + L_0^{-1} N_L~,
\label{eq:kal16}
\end{equation}
with $P_1(t) = [P(t)-P_0]/P_0$, $L_1(t) = [L(t)-L_0]/L_0$,
and $P_0=2\pi/\Omega_0$.
Equations (\ref{eq:kal13})--(\ref{eq:kal16}) are in the correct format
for analysis with a linear Kalman filter.

\section{Kalman filter
 \label{sec:kal3}}
The equations of motion (\ref{eq:kal13}) and (\ref{eq:kal14}),
and the measurement equations (\ref{eq:kal15}) and (\ref{eq:kal16}),
can be applied directly to observational data 
with the goal of estimating the model parameters using a Kalman filter.
The inputs and outputs are laid out in \S\ref{sec:kal3a},
together with a step-by-step recipe for performing the analysis.
The implementation of the parameter estimation algorithm,
which combines a Kalman filter with a nested sampler,
is outlined in \S\ref{sec:kal3b}.
The reader is referred to \citet{mey21b} for details.

\subsection{Inputs and outputs
 \label{sec:kal3a}}
The analysis takes as inputs the measured data 
as well as astrophysical priors on the model parameters.
The measurements comprise two time series, $P(t_n)$ and $L(t_n)$,
each containing $N$ samples at times $t_n$ ($1\leq n \leq N$).
The priors are left to the discretion of the analyst.
Certain model parameters appear combined inextricably as products 
in (\ref{eq:kal13})--(\ref{eq:kal16}) and cannot be estimated independently.
They comprise $M$, $R$, $I$, and one component of $(Q_0,S_0,\eta_0)$,
if the analysis follows approach I in Appendix \ref{sec:kalappb}.
They comprise $M$, $R$, and $I$ only,
if the analysis follows approach II in Appendix \ref{sec:kalappb}.
In this paper, we follow approach II, as it is more general,
and assign plausible, fiducial values to $M$, $R$, and $I$
for the sake of definiteness.

Under approach II,
the analysis returns as outputs the posteriors on the seven model parameters
${\bf\Theta}=(\gamma_\Omega,\gamma_A,\sigma_{AA})$,
with $A \in \{ Q,S,\eta \}$,
as well as an error-minimizing estimate of the time series
$\hat{\bf X}(t_n)$,
where ${\bf X} = (\Omega_1,Q_1,S_1,\eta_1)$ is the state vector.
The latter time series is potentially valuable for studying
the hidden physical processes governing
magnetospheric variables such as the Maxwell stress $S(t)$
at the disk-magnetosphere boundary.

A recipe to conduct the analysis proceeds as follows.
\begin{enumerate}
\item
Calculate $\Omega_0 = N^{-1} \sum_{n=1}^N 2\pi / P(t_n)$
and
$L_0 = N^{-1} \sum_{n=1}^N L(t_n)$.
These parameters enter the model through (\ref{eq:kal13})--(\ref{eq:kal16}).
\item
Generate the time series $P_1(t_n) = P(t_n) - 2\pi/\Omega_0$ 
and
$L_1(t_n) = L(t_n) - L_0$.
\item
Decide what model parameters to fix.
Under approach II in Appendix \ref{sec:kalappb},
one assumes plausible fiducial values for $M$, $R$, and $I$,
writes $Q_0$, $S_0$, and $\eta_0$ in terms of $\gamma_\Omega$
according to (\ref{eq:kalappb4})--(\ref{eq:kalappb6}),
and sets the parameter vector of the Kalman filter to be
${\bf\Theta}=(\gamma_\Omega, \gamma_A,\sigma_{AA})$,
with $A \in \{ Q,S,\eta \}$.
\item
Run a nested sampler like {\tt dynesty}
\citep{spe20}
with the Kalman filter likelihood defined in \S\ref{sec:kal3b}
to estimate ${\bf \Theta}$.
A formal identifiability analysis \citep{bel70} presented in Appendix \ref{sec:kalappa},
and empirical tests on synthetic data presented in \S\ref{sec:kal4},
confirm that the seven components of ${\bf \Theta}$ can be estimated unambiguously.
\item
Given ${\bf \Theta}$, estimate the time series $\hat{\bf X}(t_n)$
of the hidden state variables.
\end{enumerate}

\subsection{Implementation
 \label{sec:kal3b}}
Equations (\ref{eq:kal13})--(\ref{eq:kal16}) take the standard form of a linear Kalman filter
\citep{kal60,gel74}.
The state space representation, recursion relations, and Bayesian likelihood
of a Kalman filter are written down and justified
thoroughly in Section 3 and Appendices B--D in \citet{mey21b}.
The discussion is not repeated here in full.
The only slightly nonstandard feature of our application is that X-ray timing experiments
measure photon times of arrival, 
which must be converted to $P(t_n)$ before use.

The nested sampler evaluates the log-likelihood associated with the Kalman filter
\citep{mey21b}
\begin{equation}
 \ln p (\{ {\bf Y}_n \}_{n=1}^{N} | {\bf\Theta} )
 = 
 -\frac{1}{2} \sum_{n=1}^N
 \left[
  D_{\bf Y} \ln (2\pi) + \ln {\rm det} ( {\bf s}_n )
  + {\bf e}_n^{\rm T} {\bf s}_n^{-1} {\bf e}_n
 \right]~,
\label{eq:kal17}
\end{equation}
where ${\bf Y}_n = [P_1(t_n),L_1(t_n)]$ is the measurement vector,
$D_{\bf Y}=2$ is the dimension of ${\bf Y}_n$,
and ${\bf e}_n$ and 
${\bf s}_n = \langle {\bf e}_n {\bf e}_n^{\rm T} \rangle$ 
(Einstein summation convention suspended temporarily) 
are the innovation vector and its covariance matrix respectively.
The nested sampler proceeds iteratively.
It selects an estimate $\hat{\bf\Theta}$, runs the Kalman filter,
computes (\ref{eq:kal17}), refines $\hat{\bf\Theta}$, and repeats.
Note that ${\bf s}_n$ contains information about the measurement noise,
which is known ($\Sigma_{PP}$, $\Sigma_{LL}$), and the process noise,
which is parametrized by $\sigma_{QQ}$, $\sigma_{SS}$, and $\sigma_{\eta\eta}$ and estimated.

The innovation vector is generated at every time step $t_n$ by the Kalman filter
from the measurements and the state estimate $\hat{\bf X}_{n}$. One computes
\begin{equation}
 {\bf e}_n = {\bf Y}_n - {\bf C} \exp[{\bf A} (t_n-t_{n-1})] \hat{\bf X}_{n-1}~,
\label{eq:kal18}
\end{equation}
where ${\bf C}$ is the $2\times 4$ matrix defined implicitly 
through ${\bf Y}={\bf C}{\bf X}+(P_0^{-1} N_P,L_0^{-1} N_L)$ 
in (\ref{eq:kal15}) and (\ref{eq:kal16}).
The state vector is updated recursively via 
\begin{equation}
 \hat{\bf X}_n 
 =
 \exp[{\bf A} (t_n-t_{n-1})] \hat{\bf X}_{n-1} +{\bf k}_n {\bf e}_n~,
\label{eq:kal19}
\end{equation}
where ${\bf A}$ is the $4\times 4$ matrix in (\ref{eq:kal13})
(constant with $n$ in this application),
and ${\bf k}_n$ is the Kalman gain
defined to minimize the squared error $|{\bf X}_n - \hat{\bf X}_n|^2$.
The Kalman filter returns an estimate of the squared error at $t_n$ 
as part of its output.
An expression for the Kalman gain is provided in standard textbooks
\citep{gel74};
see also (A6) in Appendix A in \citet{kel14} 
and (C5) in Appendix C in \citet{mey21b}.

\section{Validation with synthetic data
 \label{sec:kal4}}
In this section, we lead the reader through a validation test conducted on synthetic data.
We set out the parameters of a representative test source in \S\ref{sec:kal4a}.
We then investigate the accuracy with which the Kalman filter and nested sampler 
track the state evolution and estimate the source parameters 
in \S\ref{sec:kal4b} and \S\ref{sec:kal4c} respectively.
The synthetic data are generated by solving the nonlinear equations of motion
(\ref{eq:kal5}) and (\ref{eq:kal9})--(\ref{eq:kal11})
with the Runge-Kutta It\^{o} integrator
\citep{ros10}
in the Python package {\tt sdeint}
\footnote{
{\tt https://github.com/mattja/sdeint}
}
and passing the output through the nonlinear measurement equations
(\ref{eq:kal1}) and (\ref{eq:kal2})
to produce the time series $P(t_n)$ and $L(t_n)$ for $1\leq n \leq N$.
The validation test serves two purposes:
(i) it gives an approximate sense of how accurately one can recover ${\bf \Theta}$
given a representative volume of data from a representative source;
and (ii) it is a worked example which illustrates end-to-end the practical steps
in a typical analysis.
A fuller study of the accuracy of the Kalman filter is postponed,
until the systematic (e.g.\ calibration) uncertainties associated with real astronomical data
are characterized better through collaboration with the X-ray timing community.

\subsection{Representative test source
 \label{sec:kal4a}}
As a representative example, we consider the following hypothetical source:
an accretion-powered pulsar with equilibrium mass accretion rate
$Q_0 = 3.9 \times 10^{13} \, {\rm g \, s^{-1}}$
and dipole magnetic moment
$\mu = 3.0 \times 10^{30} \, {\rm G \, cm^{3}}$.
The source parameters are recorded in Table \ref{tab:kal1}.
The components of the equilibrium state vector $(\Omega_0,Q_0,S_0,\eta_0)$
appear in  the first four lines of the table
and satisfy the conditions of magnetocentrifugal equilibrium in \S\ref{sec:kal2c}.
Observational studies of $L(t)$ and $P(t)$ fluctuations
in objects undergoing disk accretion
point to relaxation processes operating on time-scales of days to weeks
\citep{bil97,muk18,ser21},
e.g.\ the Lomb-Scargle PSD computed from the {\em RXTE} light curve of Scorpius X$-$1 
rolls over at $\sim 10^{-7} \, {\rm s^{-1}}$
\citep{muk18},
as does the PSD of $P(t_n)$ fluctuations measured in 2S 1417$-$624;
see Figure 6 in \citet{ser21}.
Hence, in the middle section of the table, we take
$1 \leq \gamma_A  / (10^{-7} \, {\rm s^{-1}}) \leq 5$
with $A \in \{ Q, S, \eta \}$
as a typical range.
\footnote{
The parameters ${\bf \Theta}$ cannot be estimated uniquely by the Kalman filter
in the special case $\gamma_Q = \gamma_\eta$, which is therefore avoided
in \S\ref{sec:kal4};
see the formal identifiability analysis in Appendix \ref{sec:kalappa}.
}
The noise amplitudes are then chosen to give fractional fluctuations of
$\lesssim 10\%$ in the hidden state variables, with
$\gamma_Q^{-1/2} \sigma_{QQ} = 0.1 Q_0$,
$\gamma_S^{-1/2} \sigma_{SS} = 0.1 S_0$,
and
$\gamma_\eta^{-1/2} \sigma_{\eta\eta} = 0.1 \eta_0$.
Fractional fluctuations of this order are broadly consistent with observations
\citep{ser21},
e.g.\ one infers $\sigma_{QQ} \approx 0.4 Q_0 \gamma_Q^{1/2}$
from Figure 3 in \citet{muk18}.
\footnote{
Translating $L(t)$ fluctuations directly to $Q(t)$ fluctuations 
via (\ref{eq:kal2}) with $\eta(t) = \eta_0 = {\rm constant}$ is an approximation,
suitable for making {\em a priori} order-of-magnitude estimates.
A self-consistent analysis involves running the Kalman filter in \S\ref{sec:kal3}
or its equivalent.
}
Finally the measurement noises $N_P(t)$ and $N_L(t)$ are chosen to be Gaussian 
for simplicity and to correspond to 1-$\sigma$ error bars of $10^{-6}$ 
and $10^{-2}$ per cent on $P_1(t_n)$ and $L_1(t_n)$ respectively.
\footnote{
$N_L(t)$ is a random error. It does not include systematic errors,
e.g.\ arising from the conversion of X-ray flux to luminosity.
In a real, astronomical analysis,
one works with $F_X(t)$ instead of $L(t)$, 
and the distance $D$ to the source cannot be inferred by the Kalman filter,
just like $M$, $R$, and $I$; see footnote \ref{foot:kal1}.
}
The error bars are deliberately conservative with respect to the brightest objects,
where the Kalman filter is likely to be tested first,
as gauged from the top-right corner of Figure 4 in \citet{ser21}.
They illustrate the most challenging parameter estimation scenario,
where the fractional amplitude of some components of the measurement noise is comparable to 
the fractional amplitude of the dynamical fluctuations in some hidden state variables.

\begin{table}[ht]
\begin{center}
\begin{tabular}{ccc}
\hline
Quantity & Value & Units \\ \hline
$\Omega_0$ & $2.2 \times 10^{-2}$ & ${\rm rad\,s^{-1}}$ \\
$Q_0$ & $3.9\times 10^{13}$ & ${\rm g\,s^{-1}}$ \\ 
$S_0$ & $9.3\times 10^0$ & ${\rm g\,cm^{-1} \, s^{-2}}$ \\
$\eta_0$ & $0.50$ & --- \\
$M$ & $2.8\times 10^{33}$ & ${\rm g}$ \\
$R$ & $1.0\times 10^6$ & ${\rm cm}$ \\
$I$ & $8.4\times 10^{44}$ & ${\rm g\, cm^2}$ \\
\hline
$\gamma_Q$ & $1.0\times 10^{-7}$ & ${\rm s^{-1}}$ \\
$\gamma_S$ & $3.0\times 10^{-7}$ & ${\rm s^{-1}}$ \\
$\gamma_\eta$ & $5.0\times 10^{-7}$ & ${\rm s^{-1}}$ \\
$\sigma_{QQ}$ & $1.3\times 10^{9}$ & ${\rm g \, s^{-3/2}}$ \\
$\sigma_{SS}$ & $5.4\times 10^{-4}$ & ${\rm g \, cm^{-1} \, s^{-5/2}}$ \\
$\sigma_{\eta\eta}$ & $3.5\times 10^{-5}$ & ${\rm s^{-1/2}}$ \\
\hline
$N$ & $500$ & --- \\
$T_{\rm obs}$ & $3.0\times 10^8$ & ${\rm s}$ \\
$\Sigma_{PP}$ & $2.9 \times 10^{-6}$ & ${\rm s}$ \\
$\Sigma_{LL}$ & $3.7 \times 10^{29}$ & ${\rm g\,cm^2 \, s^{-3}}$ \\
\hline
\end{tabular}
\end{center}
\caption{
Injected parameters of a representative accretion-powered pulsar
for the validation tests in \S\ref{sec:kal4}.
The top, middle, and bottom sections contain 
equilibrium, fluctuation, and measurement parameters respectively.
The epochs $0\leq t_1 \leq \dots \leq t_N=T_{\rm obs}$ 
are spaced equally for the sake of illustration.
Derived parameters:
$R_{{\rm m}0}= 7.3\times 10^9 \, {\rm cm}$,
$\mu= 3.0\times 10^{30} \, {\rm G\, cm^3}$,
$\gamma_\Omega=2.5\times 10^{-12} \, {\rm s^{-1}}$,
$P_0=2.9 \times 10^2 \,{\rm s}$,
$L_0 = 3.7\times 10^{33}\, {\rm g\,cm^2 \, s^{-3}}$.
}
\label{tab:kal1}
\end{table}

\subsection{State tracking
 \label{sec:kal4b}}
\begin{figure}[ht]
\begin{center}
\includegraphics[width=8cm,angle=0]{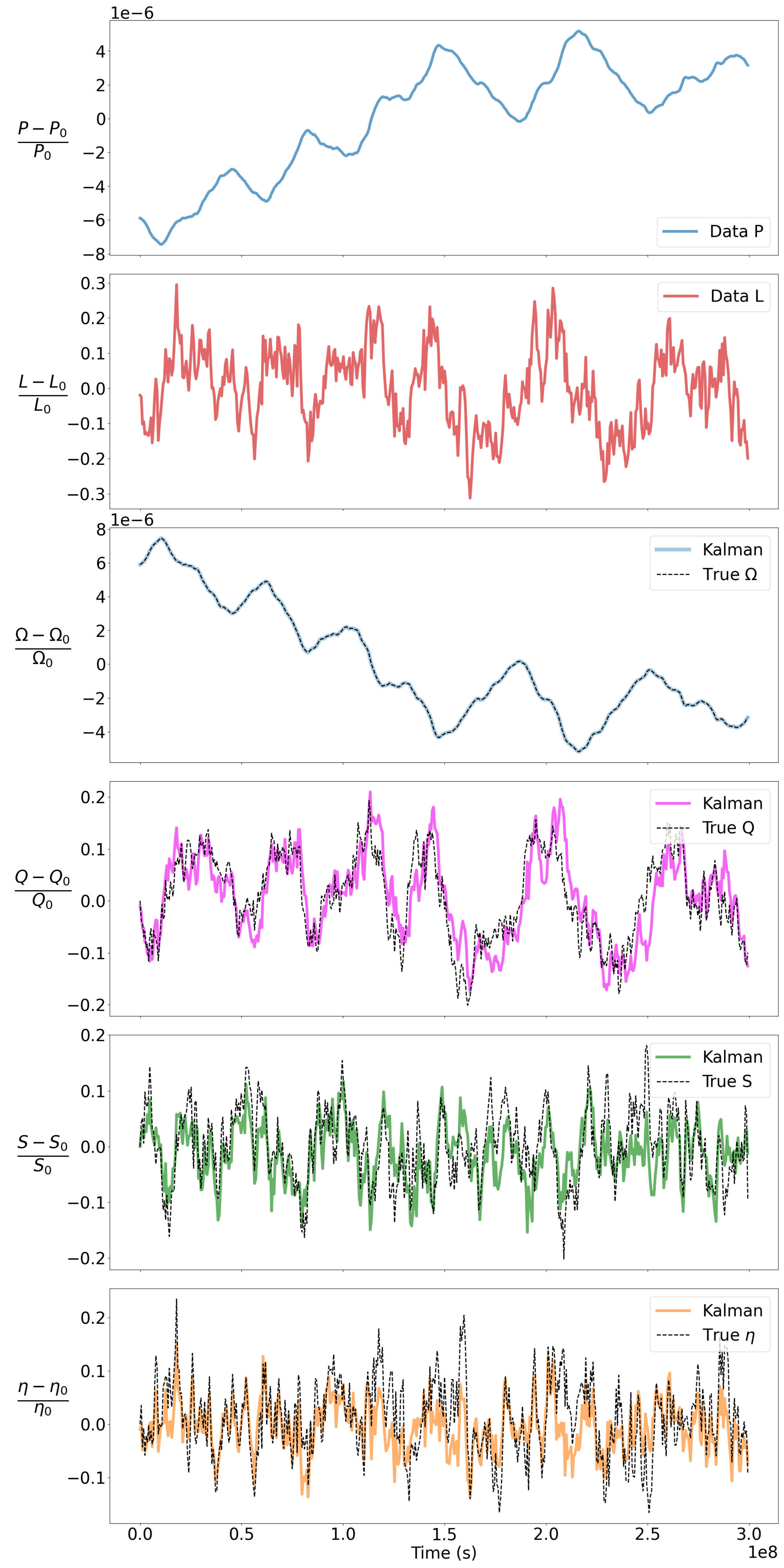}
\end{center}
\caption{
Kalman state tracking applied to the hypothetical accretion-powered pulsar 
with the parameters in Table \ref{tab:kal1}.
Inputs:
synthetic measurements of spin fluctuations $P_1(t_n)$ (top panel)
and X-ray luminosity fluctuations $L_1(t_n)$ (second panel)
versus time $t_n$ (units: ${\rm s}$), 
with $0< t_1 \leq \dots \leq t_{500} = 3.0\times 10^8\,{\rm s}$.
Outputs:
state variables
$\Omega_1(t_n)$ (third panel),
$Q_1(t_n)$ (fourth panel),
$S_1(t_n)$ (fifth panel),
and
$\eta_1(t_n)$ (bottom panel)
versus time $t_n$ (units: ${\rm s}$).
In the bottom four panels, 
colored, solid curves indicate the squared-error-minimizing,
reconstructed state sequence $\hat{\bf X}(t_n)$ generated by the Kalman filter,
and black, dashed curves indicate the true, injected state sequence ${\bf X}(t_n)$.
The vertical axis in every panel displays a fractional and therefore
dimensionless quantity.
}
\label{fig:kal1}
\end{figure}

Figure \ref{fig:kal1} presents the inputs and outputs of the Kalman filter 
as functions of time for the hypothetical source in Table \ref{tab:kal1}.
The top two panels display the synthetic measurements $P_1(t_n)$ and $L_1(t_n)$
for $1\leq n \leq 500$.
\footnote{
The sampling epochs $t_n$ are spaced equally for the sake of illustration,
but the formulas in \S\ref{sec:kal3} do not presuppose equal spacing.
}
Broadly speaking, the spin and flux wandering resemble visually and qualitatively
what one sees in real data
\citep{bil97,muk18,ser21}.
The bottom four panels display the four components of the state vector
estimated by the Kalman filter,
$\hat{\bf X}_n=[{\hat\Omega}_1(t_n),{\hat Q}_1(t_n),{\hat S}_1(t_n),{\hat \eta}_1(t_n)]$,
for $1\leq n \leq 500$.
In all four panels, there is close agreement between the estimated component
(colored, solid curve)
and the injected component generated by solving 
(\ref{eq:kal5}) and (\ref{eq:kal9})--(\ref{eq:kal11}) numerically
(black, dashed curve).
That is, the Kalman filter performs creditably in reconstructing 
the evolution of the hidden state variables $Q(t)$, $S(t)$, and $\eta(t)$,
which cannot be observed directly.
The accuracy of the reconstruction is quantified in \S\ref{sec:kal4c}.

We draw the reader's attention to two points of physical interest
in Figure \ref{fig:kal1}.
First, the mean-reverting nature of the hidden variables is clear
upon inspecting the bottom three panels.
The characteristic time-scale shortens visibly from 
$\gamma_Q^{-1} = 1.0\times 10^7 \, {\rm s}$ in the fourth panel to
$\gamma_\eta^{-1} = 2.0\times 10^6 \, {\rm s}$ in the sixth panel,
while the fractional root-mean-square amplitude 
$\approx \gamma_A^{-1/2} \sigma_{AA} = 0.1$
with $A \in \{ Q, S, \eta \}$
is the same in the fourth, fifth, and sixth panels,
as expected from Table \ref{tab:kal1}.
Second, the fractional amplitudes of the fluctuations $P_1(t_n)$
(top panel) and $\Omega_1(t_n)$ (third panel)
are $\approx \gamma_A/\gamma_\Omega \sim 10^5$ times smaller 
(with $A \in \{ Q, S, \eta \}$)
than the fluctuations in the other panels;
note the different scales on the axes.
This occurs because $\sim 10\%$ fluctuations in the torque,
driven by comparable fluctuations in the three hidden variables,
translate into smaller (and slower) fluctuations in the spin,
due to the star's large moment of inertia.
The effect is quantified analytically in Appendix \ref{sec:kalappa};
see equations (\ref{eq:kalappa6})--(\ref{eq:kalappa11}).

\subsection{Parameter estimation and its accuracy
 \label{sec:kal4c}}
\begin{figure}[ht]
\begin{center}
\includegraphics[width=15cm,angle=0]{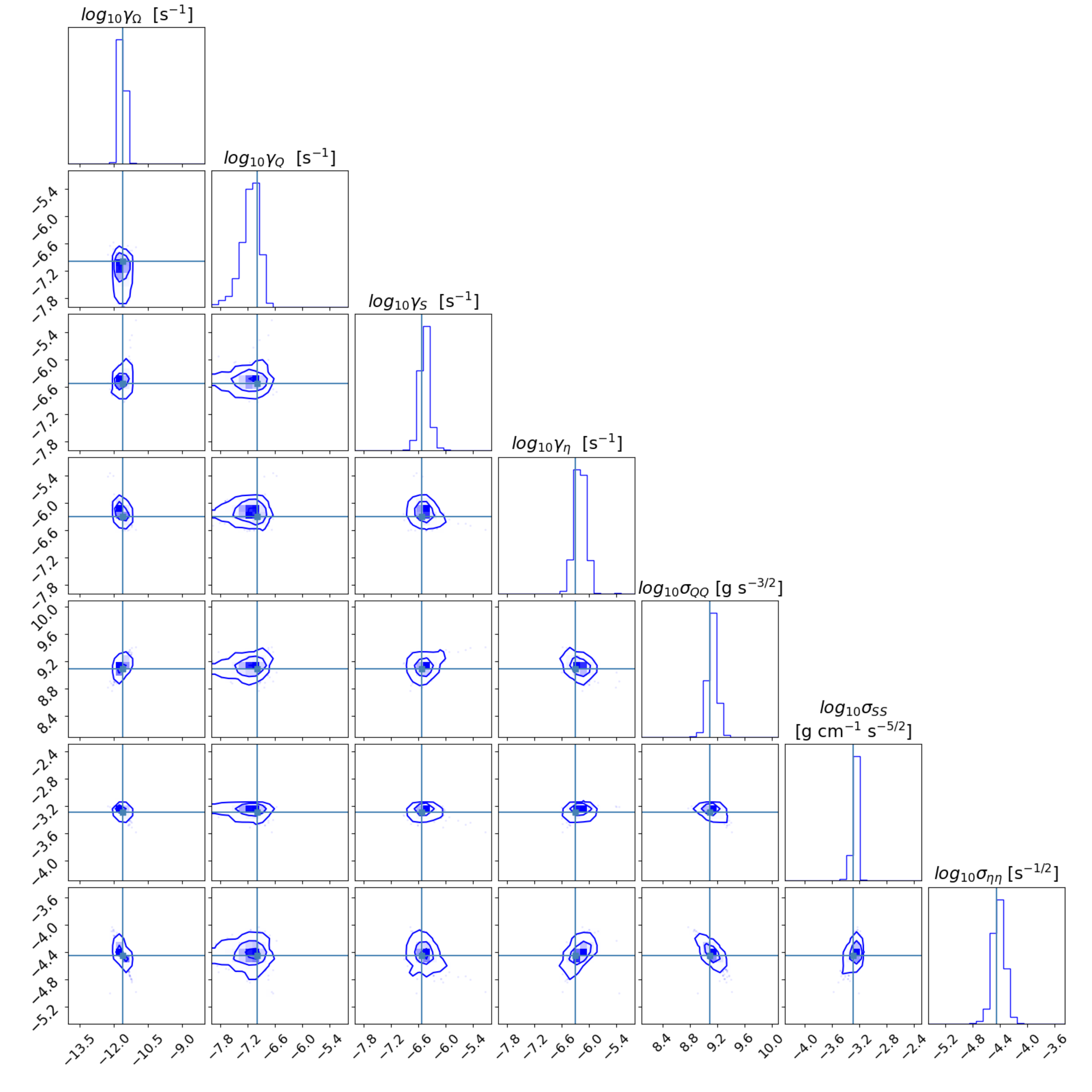}
\end{center}
\caption{
Corner plot of the posterior distribution of the model parameters ${\bf\Theta}$
for the hypothetical accretion-powered pulsar in Table \ref{tab:kal1},
viz.\ 
$\gamma_\Omega$, $\gamma_Q$, $\gamma_S$, $\gamma_\eta$,
$\sigma_{QQ}$, $\sigma_{SS}$, and $\sigma_{\eta\eta}$ 
(left to right and top to bottom).
All quantities are plotted on a log scale (base 10).
Contour plots depict the posterior distribution marginalized over five out of seven parameters, 
e.g.\ the bottom left corner displays the marginalized posterior 
in the $\gamma_\Omega$-$\sigma_{\eta\eta}$ plane.
Histograms depict the posterior marginalized over six out of seven parameters.
The injected parameter values are marked by horizontal and vertical blue lines.
The blue lines coincide approximately with the modes of the marginalized posteriors in every panel,
implying that the parameters are estimated accurately.
}
\label{fig:kal2}
\end{figure}

Figure \ref{fig:kal2} presents the posterior distribution of the seven parameters
${\bf \Theta} = (\gamma_\Omega,\gamma_A,\sigma_{AA})$,
with $A \in \{ Q, S, \eta \}$,
returned by the nested sampler in \S\ref{sec:kal3}.
The mode of the posterior corresponds to
the optimal estimated state sequence $\hat{\bf X}_n$ plotted in Figure \ref{fig:kal1}.
The seven-dimensional posterior is visualized in cross-section
through a traditional corner plot.
All seven parameters are estimated unambiguously and accurately.
This empirical finding confirms the prediction of a formal identifiability analysis,
a standard tool in electrical engineering
\citep{bel70},
which is presented in Appendix \ref{sec:kalappa}.
\footnote{
In many systems, certain parameters cannot be estimated unambiguously.
For example, radio timing data are insufficient in general
to identify all six parameters of the classic, two-component, crust-superfluid model 
of a pulsar exhibiting timing noise;
see \citet{mey21b}.
}
The one-dimensional posteriors (histograms) in Figure \ref{fig:kal2} are unimodal
and peak near the injected parameter values indicated by the blue lines,
with the absolute error in the peak ranging from a minimum of 
$\approx 0.025\,{\rm dex}$ for $\gamma_Q$
to a maximum of $\approx 0.15 \, {\rm dex}$ for $\gamma_\Omega$.
Likewise the two-dimensional posteriors (contour plots) are unimodal
and peak near the injection (intersection of the blue lines).
The full width half maximum ranges from $\approx 0.10\,{\rm dex}$ for $\sigma_{QQ}$
and $\sigma_{SS}$
to $\approx 0.60\,{\rm dex}$ for $\gamma_{\Omega}$.
The parameters in ${\bf \Theta}$ are largely uncorrelated.
Arguably there are hints of $\gamma_\Omega$-$\sigma_{\eta\eta}$
and $\sigma_{QQ}$-$\sigma_{\eta\eta}$ anticorrelations,
visible as a diagonal tilt of the innermost contours 
in the $\gamma_\Omega$-$\sigma_{\eta\eta}$ and $\sigma_{QQ}$-$\sigma_{\eta\eta}$ planes,
but their significance is marginal.

\begin{figure}[ht]
\begin{center}
\includegraphics[width=15cm,angle=0]{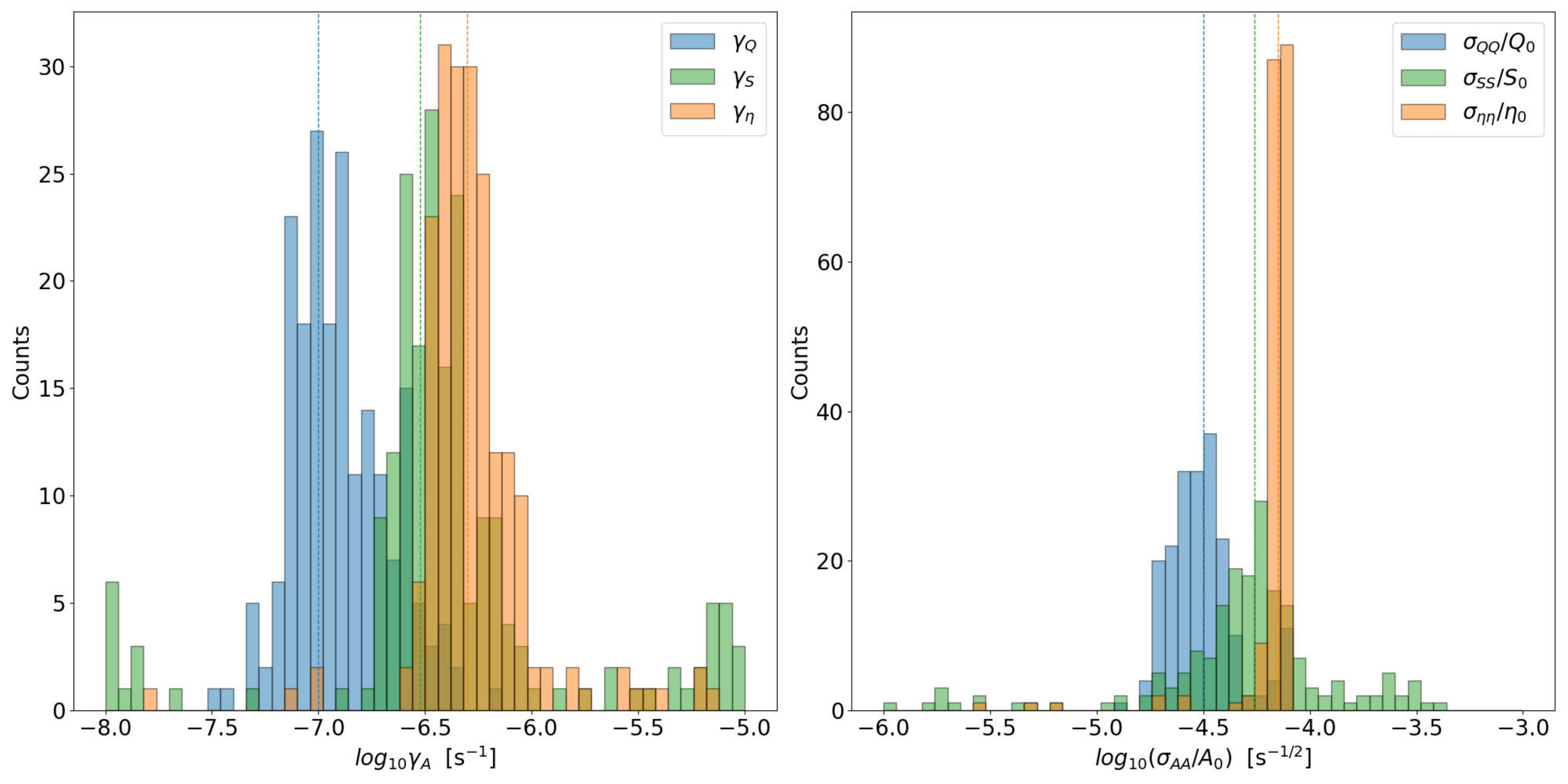}
\end{center}
\caption{
Accuracy of parameter estimation with the Kalman filter and nested sampler
for the hypothetical accretion-powered pulsar in Table \ref{tab:kal1}.
(Left panel.) 
Histograms of estimated $\gamma_A$ values (units: ${\rm s^{-1}}$),
with $A=Q$ (blue), $S$ (green), and $\eta$ (orange),
recovered from 200 random realizations of the synthetic data $P(t_n)$ and $L(t_n)$.
The dashed, color-coded, vertical lines correspond to the injected $\gamma_A$ values
in Table \ref{tab:kal1}.
All $3\times 200$ estimates fall within the plotted domain.
(Right panel.)
Histograms of estimated $\sigma_{AA}$ values,
normalized by the equilibrium value $A_0$ for ease of display,
with $A=Q$ (blue; units of ${\rm s^{-1/2}}$), 
$S$ (green; units of ${\rm s^{-1/2}}$), 
and $\eta$ (orange; units of ${\rm s^{-1/2}}$).
The dashed, color-coded, vertical lines correspond to the injected values.
All estimates fall within the plotted domain,
except for 13 $\sigma_{SS}$ and five $\sigma_{\eta\eta}$ outliers,
whose logarithms satisfy $-8 \leq \log_{10} (\sigma_{AA} / A_0) \leq -6$.
}
\label{fig:kal3}
\end{figure}

The results in Figures \ref{fig:kal1} and \ref{fig:kal2} 
refer to a single, random realization of the noisy measurements
$P(t_n)$ and $L(t_n)$. How representative are they, if the experiment is repeated?
Figure \ref{fig:kal3} demonstrates how accurately the Kalman filter and nested sampler
recover the injected values of the fluctuation parameters 
$\gamma_A$ (left panel) and $\sigma_{AA}/A_0$ 
(right panel; normalized by the equilibrium value $A_0$)
in Table \ref{tab:kal1},
for $A \in \{ Q, S, \eta \}$ color-coded as per the caption,
by analysing 200 random realizations of the noisy measurements.
Each variable is plotted in cgs units on a log scale.
All six histograms peak near the injected values,
indicated by dashed, color-coded, vertical lines.
The absolute error in the peak ranges from $\approx 0.010 \, {\rm dex}$ for $\gamma_Q$
to $\approx  0.11 \, {\rm dex}$ for $\gamma_{\eta}$.
Moreover the dispersion is modest.
The full width half maximum ranges from
$\approx 0.12 \,{\rm dex}$ for $\sigma_{\eta\eta}$
to $\approx 0.60 \, {\rm dex}$ for $\gamma_Q$.
Extreme outliers (e.g.\ when the nested sampler fails to converge) are rare.
All $3\times 200$ $\gamma_A$ estimates lie within the plotted domain.
The same holds for $\sigma_{AA}$,
except for 13 $\sigma_{SS}$ and five $\sigma_{\eta\eta}$ estimates,
which satisfy $-8 \leq \log_{10} (\sigma_{AA}/A_0) \leq -6$.
Similarly encouraging results are obtained for $\gamma_\Omega$ 
and are not overplotted for the sake of readability.

\section{Conclusion
 \label{sec:kal5}}
The accretion disk and magnetosphere of an accretion-powered pulsar
form a stochastic dynamical system, 
driven by complicated processes such as flicker noise in the disk $\alpha$ parameter
and self-healing Rayleigh-Taylor instabilities at the disk-magnetosphere boundary.
Simultaneous, high-time-resolution measurements of the aperiodic X-ray luminosity $L(t)$
and spin period $P(t)$ can be combined to probe the disk-magnetosphere physics.
Most previous, pioneering work in this direction involves temporal averaging
explicitly or implicitly,
e.g.\ when measuring and interpreting the scaling between flux and root-mean-square variability
\citep{rev15,pat21},
the scaling between torque and luminosity
\citep{bay93,san17,ser21},
and the torque and luminosity PSDs
\citep{bil97,rig08,rev09,pat21}.
However the specific, time-ordered sequence of noisy measurements $L(t_n)$ and $P(t_n)$
contains a great deal of useful, additional information, 
such as instantaneous (anti)correlations between the system variables,
which is lost after averaging over time.

In this paper, we show how to extract the instantaneous information in the
stochastic time series $L(t_n)$ and $P(t_n)$
using a standard Kalman filter with the discrete-time structure 
set out in \S\ref{sec:kal3}.
We show that it is possible in principle to measure important accretion parameters,
which cannot be disentangled by a time-averaged analysis.
The Kalman filter relates $L(t_n)$ and $P(t_n)$ to three hidden state variables,
which are of prime physical interest but cannot be measured directly:
the mass accretion rate $Q$, the Maxwell stress $S$ at the disk-magnetosphere boundary,
and the radiative efficiency $\eta$.
The state variables evolve
according to the canonical magnetocentrifugal theory of accretion,
linearized about magnetocentrifugal equilibrium,
and execute mean-reverting random walks driven by white noise.
Colored noise and modified forms of mean reversion
fit comfortably within the formalism, if future data demand their inclusion;
preliminary extensions to magnetocentrifugal disequilibrium and a generalized accretion model 
are outlined briefly in Appendices \ref{sec:kalappc} and \ref{sec:kalappd} respectively
by way of illustration.
The linear Kalman filter is defined completely by the dynamical equations
(\ref{eq:kal13}) and (\ref{eq:kal14}),
the measurement equations
(\ref{eq:kal15}) and (\ref{eq:kal16}),
and the equilibrium state $(\Omega_0, Q_0, S_0, \eta_0)$,
which is determined from the time-averaged and Kalman filtered data
assuming fiducial values of $M$, $R$, and $I$.

Tests on synthetic data in \S\ref{sec:kal4} demonstrate that the Kalman filter,
combined with a nested sampler,
recovers the seven model parameters ${\bf \Theta} = (\gamma_\Omega, \gamma_A, \sigma_{AA})$
with $A\in \{ Q, S, \eta \}$ unambiguously,
confirming the formal identifiability analysis in Appendix \ref{sec:kalappa}.
It achieves an ensemble-averaged accuracy of better than $\approx 0.1\,{\rm dex}$
(absolute error in the mode)
and a dispersion of less than $\approx 0.6 \, {\rm dex}$ 
(full width half maximum)
for every ${\bf \Theta}$ component,
assuming typical data volumes,
viz.\ a few hundred measurements,
and parameters representative of accretion-powered pulsars.
Interestingly, the method estimates not just the fluctuation parameters
$\gamma_A$ and $\sigma_{AA}$ but also the equilibrium state through $\gamma_\Omega$,
breaking the degeneracy that exists between $\Omega_0$, $Q_0$, $S_0$, and $\eta_0$,
when one is restricted to time-averaged data only.
By doing so, the method makes it possible to measure the star's magnetic moment $\mu$,
at least in principle, which is important physically.
The tests in \S\ref{sec:kal4} serve as a practical, end-to-end tutorial
on how to apply the method, from computing the equilibrium state
to running the Kalman filter and interpreting its output.

The next step is to apply the Kalman filter to real,
astronomical data in collaboration with the X-ray timing community.
As just one example,
it would be interesting to reanalyse the source 2S 1417$-$624,
a bright, transient, accretion-powered pulsar
\citep{ser21}.
At the time of writing,
time series of $P(t_n)$ and the pulsed X-ray flux --- but not $L(t_n)$ ---
have been released publicly for 2S 1417$-$624
by the {\em Fermi} Gamma-Ray Burst Monitor (GBM)
Accreting Pulsars Program.
\footnote{
{\tt https://gammaray.msfc.nasa.gov/gbm/science/pulsars.html}
}
The same source was observed independently by {\em NICER} during its 2018 outburst
in three intervals between MJD 58211 and MJD 58350.
In the {\em NICER} data,
$L(t_n)$ is available publicly but $P(t_n)$ is not,
the opposite of the situation with {\em Fermi}.
One can generate $P(t_n)$ from public data in principle,
but the timing analysis is not trivial and needs involvement from experts
with custom-designed software
and a thorough understanding of the systematic (e.g.\ calibration) uncertainties;
see {\S}4.1 in \citet{ser21} for details.
Looking ahead to the future,
it would be advantageous to release simultaneous $P(t_n)$ and $L(t_n)$ time series
for every target in the {\em Fermi} GBM Accreting Pulsars Program.
With these data,
it should be possible to compile statistics
about the fluctuation parameters $\gamma_A$ and $\sigma_{AA}$
with $A\in \{ Q, S, \eta \}$
across the subset of the accreting pulsar population,
where the canonical magnetocentrifugal model is thought to describe
the accretion physics to a reasonable approximation.

A promising avenue for future work is gravitational wave astronomy.
Spin wandering is a key limitation on searches
for continuous gravitational radiation from low-mass X-ray binaries
\citep{wat08,ril13,muk18,ril22}.
The coherence time $T_{\rm drift}$ of a search with an optimal matched filter,
such as the maximum likelihood ${\cal F}$-statistic
\citep{jar98},
must be short enough,
such that the quasimonochromatic signal stays inside a single Fourier bin
throughout the coherent integration.
That is, its root-mean-square frequency fluctuation must satisfy
\citep{suv17}
\begin{equation}
 \langle \delta\Omega(T_{\rm drift})^2 \rangle^{1/2}
 \leq 
 \pi / T_{\rm drift}~.
\label{eq:kal20}
\end{equation}
In nonpulsating systems, such as the high-priority LIGO target Scorpius X$-$1,
where $\Omega(t)$ cannot be measured directly,
one is obliged to guess $T_{\rm drift}$ by analogy from the PSD of $L(t)$;
a popular choice is $T_{\rm drift} \sim 10\,{\rm days}$ for Scorpius X$-$1
\citep{muk18}.
An improved theoretical understanding of the connection between $\Omega(t)$
and $L(t)$ in magnetocentrifugal accretion,
facilitated by the Kalman filter in this paper,
would help refine PSD-based estimates of $T_{\rm drift}$.
Indeed, it may even be possible
to estimate directly a subset of the parameters ${\bf \Theta}$
with a Kalman filter from the time series $L(t_n)$ alone,
as in other neutron star applications where observations are incomplete
\citep{mey21b};
see also \citet{mey21a}.
Specifically, if it is possible to estimate
$\gamma_\Omega$, $\gamma_Q$, $\gamma_S$, $\sigma_{QQ}$, and $\sigma_{SS}$ from $L(t_n)$,
one can solve the inequality (\ref{eq:kal20}) for the maximum $T_{\rm drift}$
by substituting (\ref{eq:kalappa6}) into the left-hand side of (\ref{eq:kal20}).
We will explore this opportunity in a forthcoming paper.

\acknowledgments
The authors thank Katie Auchettl for discussions about X-ray observations
of accretion-powered pulsars.
We are also grateful to the anonymous referee for specific feedback 
and concrete suggestions about the idealizations in the accretion physics, 
which clarified several important points and
inspired the calculations in Appendices \ref{sec:kalappc} and \ref{sec:kalappd}.
This research was supported by the Australian Research Council
Centre of Excellence for Gravitational Wave Discovery (OzGrav),
grant number CE170100004.
NJO'N is the recipient of a Melbourne Research Scholarship.

\bibliographystyle{mn2e}
\bibliography{amsp}

\appendix
\section{Equilibrium state vector in terms of observables
 \label{sec:kalappb}}
The magnetocentrifugal equilibrium defined in \S\ref{sec:kal2c}
is described by the state vector $(\Omega_0,Q_0,S_0,\eta_0)$,
whose four components are formal inputs into the Kalman filter in \S\ref{sec:kal3}.
In this appendix it is shown how to solve for $(\Omega_0,Q_0,S_0,\eta_0)$
following approaches I and II introduced in \S\ref{sec:kal2c},
which use time-averaged and Kalman filtered data respectively.
In what follows, we assume that $M$, $R$, and $I$ are known {\em a priori}
on theoretical grounds.
They occur in inseparable combinations and cannot be inferred uniquely from the data.

Upon averaging the measurements $P(t_n)$ and $L(t_n)$ over time, 
and combining with the condition for magnetocentrifugal equilibrium,
we obtain three pieces
of independent information about $(\Omega_0,Q_0,S_0,\eta_0)$.
From the definitions of $\Omega_0$ and $L_0$ in \S\ref{sec:kal2c},
we have
\begin{equation}
 \Omega_0 = N^{-1} \sum_{n=1}^N {2\pi} / {P(t_n)}
\label{eq:kalappb1}
\end{equation}
and
\begin{equation}
 L_0 = N^{-1} \sum_{n=1}^N L(t_n)~.
\label{eq:kalappb2}
\end{equation}
Equation (\ref{eq:kalappb1}) gives $\Omega_0$ directly.
Secondly, combining (\ref{eq:kalappb1}) with the zero-torque condition (\ref{eq:kal6})
implies a relation between $Q_0$ and $S_0$.
Thirdly, combining (\ref{eq:kalappb2}) with the energy conservation law (\ref{eq:kal7})
implies a relation between $Q_0$ and $\eta_0$.
The sample means (\ref{eq:kalappb1}) and (\ref{eq:kalappb2})
do not equate exactly to the true equilibrium values of the dynamical variables $\Omega(t)$ and $L(t)$
in (\ref{eq:kal6}) and (\ref{eq:kal7}) due to random dispersion of fractional order $N^{-1/2}$
and systematic errors if the accretion physics is not stationary.
However, there is no realistic alternative to using the sample means;
it is a standard approach when interpreting data in terms of the 
magnetocentrifugal accretion paradigm
\citep{pat21}.

The two relations between $Q_0$, $S_0$ and $\eta_0$ obtained from
(\ref{eq:kal6}) and (\ref{eq:kal7}),
using the data in (\ref{eq:kalappb1}) and (\ref{eq:kalappb2}),
must be supplemented by another piece of information to solve for
$Q_0$, $S_0$ and $\eta_0$ uniquely.
Approach I in \S\ref{sec:kal2c} involves assuming
a plausible value for one unknown,
solving for the other two,
and running the Kalman filter in \S\ref{sec:kal3} 
with all four components of $(\Omega_0,Q_0,S_0,\eta_0)$ fixed.
It is natural but not obligatory to assume $\eta_0$,
which is bounded ($0 < \eta_0 < 1$)
and widely believed to satisfy $\eta_0 \sim 1$
\citep{bil97,san17}.

Approach II in \S\ref{sec:kal2c} is more general:
it keeps one component of $(\Omega_0,Q_0,S_0,\eta_0)$ free to be estimated.
Specifically, the linearized equations of motion (\ref{eq:kal13})
feature the parameter 
$\gamma_\Omega[\Omega_0,Q_0,R_{{\rm m}0}(\Omega_0)]$
defined by (\ref{eq:kal14}), viz.\
\begin{equation}
 \gamma_\Omega
 =
 \frac{(GM)^{2/3} Q_0}{I \Omega_0^{4/3}}~,
\label{eq:kalappb3}
\end{equation}
where (\ref{eq:kalappb3}) follows from (\ref{eq:kal14}) and
$R_{{\rm m}0} = R_{{\rm c}0} = (GM)^{1/3} \Omega_0^{-2/3}$.
Approach II involves running the parameter estimation scheme with $\gamma_\Omega$ free,
so that it is one of the parameters estimated from the data.
This yields a relation between $\Omega_0$ and $Q_0$,
which is combined with (\ref{eq:kalappb1}) and the relations derived
from (\ref{eq:kal6}) and (\ref{eq:kal7}) to solve for all four components
of $(\Omega_0,Q_0,S_0,\eta_0)$.
Handy formulas are quoted below for the convenience of the reader:
\begin{eqnarray}
 Q_0 
 & = &
 \frac{I \Omega_0^{4/3} \gamma_\Omega}{(GM)^{2/3}}~,
\label{eq:kalappb4}
 \\
 S_0 
 & = &
 \frac{I \Omega_0^3 \gamma_\Omega}{4\pi GM}~,
\label{eq:kalappb5}
 \\
 \eta_0
 & = &
 \frac{R L_0}{(GM)^{1/3} I \Omega_0^{4/3} \gamma_\Omega}~.
\label{eq:kalappb6}
\end{eqnarray}
We reiterate that $M$, $R$, and $I$ cannot be inferred uniquely from the data;
fiducial values are inserted instead.

\section{Magnetocentrifugal disequilibrium
 \label{sec:kalappc}}

\subsection{Intervals of secular acceleration and deceleration
 \label{sec:kalappca}}
Many accretion-powered pulsars exist well away from the state of 
magnetocentrifugal equilibrium,
satisfying $| R_{\rm c}(t) - R_{\rm m}(t) | \gtrsim R_{\rm c}(t)$ for example
\citep{bil97,yan17,mus22,ser22}.
Disequilibrium occurs in low-mass X-ray binaries, 
such as Her X$-$1 and 4U 1626$-$67,
which accrete via a persistent disk fed by Roche lobe overflow
\citep{gen22};
symbiotic X-ray binaries, such as GX 1$+$4,
which may involve quasispherical accretion
\citep{gon12};
and high-mass X-ray binaries, 
such as GX 301$-$2, 4U 1538$-$52, OAO 1657$-$415, Vela X$-$1, and Cen X$-$3,
which accrete via a transient disk fed by a supergiant wind
(and possibly Roche lobe overflow too)
\citep{ser22}.
In disequilibrium, 
the star alternates between long spin-up and spin-down episodes,
which last for $\gtrsim 10\,{\rm yr}$,
separated by fast transitions lasting as short as $\sim 1 \, {\rm day}$,
e.g.\ in 4U 0115$+$63
\citep{cam01};
see Figures 6 and 31 in \citet{bil97},
Figure 19 in \citet{mus22},
and Figures 1--7 in \citet{ser22} for typical examples.
Sometimes, as in A 0535$+$26, X-ray outbursts and quiescence are correlated 
with acceleration and deceleration episodes respectively;
see Figure 19 in \citet{mus22}.

The Kalman filter technique introduced in this paper is new in the context of
parameter estimation for accretion-powered pulsars.
It is prudent, therefore, to validate it first 
with objects that are near magnetocentrifugal equilibrium,
where the linear theory in \S\ref{sec:kal2e} and 
linear Kalman filter in \S\ref{sec:kal3} apply without modification,
before attempting the more ambitious application to disequilibrium.
Real data exist on many objects near equilibrium,
e.g.\ 30 out of the 65 known X-ray pulsars in the Small Magellanic Cloud,
labeled by the letter `C' in Table 3 in \citet{yan17},
whose time-averaged $dP/dt$ values are smaller than 1.5 times the
measurement error on $dP/dt$.

It is straightforward in principle to generalize the analysis 
in \S\ref{sec:kal2} and \S\ref{sec:kal3} to objects in disequilibrium,
by analyzing the spin-up and spin-down intervals separately.
An advantage in doing so is that one can compare and gain physical insight from
the parameter values estimated independently from consecutive intervals.
Some parameters (e.g.\ $\eta_0$, if dominated by radiative processes)
may be expected to remain roughly unchanged,
whereas others (e.g.\ $Q_0$, if driven by the 
companion star's mass loss rate) may change substantially.
Another advantage is that an interval-by-interval analysis does not rely on
modifying (\ref{eq:kal5}) and (\ref{eq:kal9})--(\ref{eq:kal11}) to track the
fast torque transition,
which is not understood fully yet in terms of simple analytic laws
like (\ref{eq:kal5}) and (\ref{eq:kal9})--(\ref{eq:kal11}),
which are needed to implement a Kalman filter
\citep{nel97,van98,loc04,lai14,gen22}.
The torque transition is related to the challenging physics of the disk-magnetosphere boundary,
which is surveyed briefly in Appendix \ref{sec:kalappd}.

\subsection{Worked example: unscented Kalman filter applied to a secular interval
 \label{sec:kalappcb}}
We present for the sake of completeness a short worked example to illustrate
how a simple, nonlinear adjustment of the framework in \S\ref{sec:kal2} 
can be made to analyze
an individual spin-up or spin-down interval satisfying $R_{\rm c}(t) \neq R_{\rm m}(t)$.
The adjustment proceeds in two steps.
First, we replace the linearized measurement equations (\ref{eq:kal15}) and (\ref{eq:kal16})
with their nonlinear counterparts (\ref{eq:kal1}) and (\ref{eq:kal2}) respectively.
Second, we replace the linearized equations of motion (\ref{eq:kal13}) and (\ref{eq:kal14})
with the nonlinear torque law (\ref{eq:kal5}) and new Langevin equations,
\begin{eqnarray}
 \frac{dQ}{dt}
 & = &
 -\gamma_Q [Q(t) - \overline{Q} ] + \xi_Q(t)~,
\label{eq:kalappc1}
 \\
 \frac{dS}{dt}
 & = &
 -\gamma_S [S(t) - \overline{S} ] + \xi_S(t)~,
\label{eq:kalappc2}
 \\
 \frac{d\eta}{dt}
 & = &
 -\gamma_\eta [\eta(t) - \overline{\eta} ] + \xi_\eta(t)~.
\label{eq:kalappc3}
\end{eqnarray}
Equations (\ref{eq:kalappc1})--(\ref{eq:kalappc3}) describe linear mean reversion
in common with (\ref{eq:kal9})--(\ref{eq:kal11});
see Appendix \ref{sec:kalappd} for some of the refinements required to make the 
mean reversion more realistic physically through nonlinear feedback.
Indeed, equations (\ref{eq:kalappc1})--(\ref{eq:kalappc3})
are identical mathematically to (\ref{eq:kal9})--(\ref{eq:kal11}).
However, they are interpreted differently.
For example,
$\overline{Q} = \langle Q(t) \rangle$ in (\ref{eq:kalappc1}),
the ensemble average of $Q(t)$,
replaces the equilibrium quantity $Q_0$ in (\ref{eq:kal9}).
Taking $\overline{Q} > Q_0$,
we have $\overline{R}_{\rm m} = \langle R_{\rm m}(t) \rangle < R_{\rm c}$
and hence $d\Omega / dt > 0$ on average;
the magnetosphere is compressed relative to equilibrium.
In contrast, 
$\overline{Q} < Q_0$ implies $d\Omega / dt < 0$, i.e.\ the propeller phase.
Here $\overline{Q}$ is a static parameter to be estimated by the Kalman filter.

Kalman parameter estimation for nonlinear problems,
such as (\ref{eq:kal1}), (\ref{eq:kal2}), (\ref{eq:kal5}), 
and (\ref{eq:kalappc1})--(\ref{eq:kalappc3}),
is a standard procedure.
Algorithms include the extended and unscented Kalman filters
\citep{jaz70,jul97}.
In the worked example in this appendix,
we use an unscented Kalman filter,
which shares the same predictor-corrector design as the linear Kalman filter.
It works on the principle that a set of weighted sampling points (`sigma points'),
generated deterministically from the matrix square root of the state covariance matrix,
are used to estimate the mean and covariance of the conditional probability density.
The algorithm is laid out in standard textbooks and in Box 3.1 of the original paper
by \citet{jul97}.
The Kalman log-likelihood is given by (\ref{eq:kal17}) and evaluated by the
{\tt dynesty} nested sampler \citep{spe20}.
\footnote{
In an unscented Kalman filter,
${\bf s}_n = \langle {\bf e}_n {\bf e}_n^{\rm T} \rangle$
in (\ref{eq:kal17}) is often called the predicted measurement covariance
and is calculated as a weighted average over the sigma points
\citep{jul97}.
}
The parameters to be estimated are 
${\bf\Theta} = (\beta_1, \beta_2, \gamma_A, \sigma_{AA} )$, 
with 
\begin{equation}
 \beta_1 
 =
 \frac{(GM)^{3/5} \overline{Q}^{6/5}}
  {(4\pi)^{1/5} I \overline{S}^{1/5} \overline{\Omega} }~,
\label{eq:kalappc4}
\end{equation}
\begin{equation}
 \beta_2
 =
 \frac{(GM)^{2/5} \overline{Q}^{9/5} }
  {(4\pi)^{4/5} I \overline{S}^{4/5} }~,
\label{eq:kalappc5}
\end{equation}
and $A \in \{ Q, S, \eta \}$,
i.e.\ eight parameters instead of the seven in the linearized recipe in \S\ref{sec:kal3}.
In (\ref{eq:kalappc4}),
$\overline{\Omega} = \langle \Omega(t) \rangle$ is obtained directly from the data
by calculating the sample average (\ref{eq:kalappb1})
without assuming an attendant equilibrium interpretation.
Once ${\bf\Theta} = (\beta_1, \beta_2, \gamma_A, \sigma_{AA} )$ is estimated,
we solve (\ref{eq:kalappc4}) and (\ref{eq:kalappc5})
for $\overline{Q}$ and $\overline{S}$ and then solve
\begin{equation}
 \overline{L} = GM \overline{Q \eta} / R
\label{eq:kalappc6}
\end{equation}
for $\overline{\eta}$,
where $\overline{L} = \langle L(t) \rangle$ is obtained directly from the data 
by calculating the sample average (\ref{eq:kalappb2}),
again without assuming equilibrium.
In this appendix, for the sake of simplicity, we assume that $\overline{Q\eta}$
reduces to $\overline{Q}$ times $\overline{\eta}$,
as we do not estimate $\overline{\eta}$ in the worked example below. 
In general, however, $Q(t)$ and $\eta(t)$ are correlated, 
and 
\begin{equation}
 \beta_3 
 =
 \frac{ \langle Q(t) \eta(t) \rangle }{ \langle Q(t) \rangle \langle \eta(t) \rangle }
\label{eq:kalappc7}
\end{equation}
represents a ninth parameter to be appended to ${\bf\Theta}$ and estimated.

\begin{figure}[ht]
\begin{center}
\includegraphics[width=11cm,angle=0]{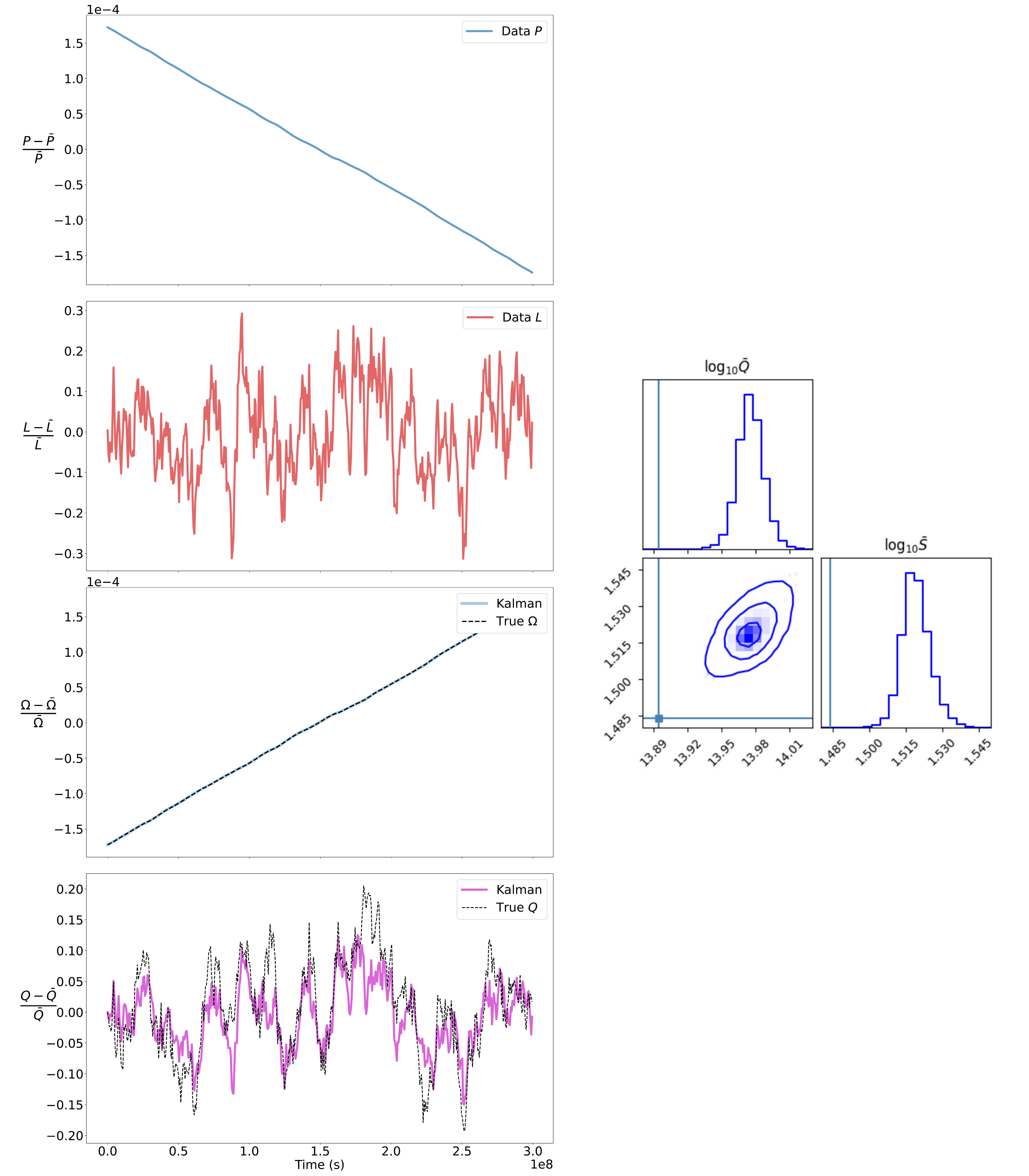}
\end{center}
\caption{
Kalman state tracking applied to a hypothetical accretion-powered pulsar 
in disequilibrium
with the parameters in Table \ref{tab:kal1}
except for
$\overline{Q} = 2 Q_0$ and $\overline{S} = 2^{12/7} S_0$, i.e.\ spinning up.
Inputs:
synthetic measurements of spin period $P(t_n)$ (top panel at left)
and X-ray luminosity $L(t_n)$ (second panel at left)
versus time $t_n$ (units: ${\rm s}$), 
with $0< t_1 \leq \dots \leq t_{500} = 3.0\times 10^8\,{\rm s}$.
Outputs:
state variables
$\Omega(t_n)$ (third panel at left)
and
$Q(t_n)$ (fourth panel at left)
versus time $t_n$ (units: ${\rm s}$);
the other state variables $S(t_n)$ and $\eta(t_n)$
are omitted for clarity, as they resemble Figure \ref{fig:kal1}.
Parameter estimation:
posterior of $\overline{Q}$ and $\overline{S}$,
marginalized over $(\gamma_A,\sigma_{AA})$ with $A \in \{ Q, S, \eta \}$
and presented as a traditional corner plot
(contour plot and histograms; panels at right).
The injected values are marked by blue horizontal and vertical lines.
In the third and fourth panels at left,
colored, solid curves indicate the squared-error-minimizing,
reconstructed state sequence $\hat{\bf X}(t_n)$ generated by the Kalman filter,
and black, dashed curves indicate the true, injected state sequence ${\bf X}(t_n)$.
The vertical axes in the four panels at left display fractional and therefore
dimensionless quantities.
}
\label{fig:kalappc1}
\end{figure}

Figure \ref{fig:kalappc1} displays the tracking and estimation results
for a hypothetical pulsar in disequilibrium.
Its injected parameters are the same as in Table \ref{tab:kal1},
but we take $\overline{Q} = 2 Q_0$ in (\ref{eq:kalappc1}),
so that the pulsar spins up secularly throughout the observation.
(Secular spin down, e.g.\ $\overline{Q} = 0.5 Q_0$, works the same way.)
We also take $\overline{S} = 2^{12/7} S_0$ in (\ref{eq:kalappc2})
to keep $\mu \propto \overline{S}^{-7/10} \overline{Q}^{6/5}$ unchanged
relative to Table \ref{tab:kal1}.
One sees clearly the expected downward trend in the synthetic measurements of $P(t_n)$
in the top panel on the left,
and the upward trend in the tracked state variable $\Omega(t_n)$
in the third panel on the left.
One also sees $Q(t)$ executing mean-reverting fluctuations around $\overline{Q}$
in the fourth panel on the left, in line with (\ref{eq:kalappc1}).
By inspection, the Kalman state estimates (colored curves) 
track the injected state sequence (black, dashed curves) accurately.
The synthetic measurements of $L(t_n)$ resemble qualitatively those in Figure \ref{fig:kal1},
as do the Kalman state estimates of $S(t)$ and $\eta(t)$ (not plotted for clarity).

The right panel of Figure \ref{fig:kalappc1}
displays the posterior distribution (corner plot) of $\overline{Q}$ and $\overline{S}$,
inferred from $\beta_1$ and $\beta_2$ via (\ref{eq:kalappc4}) and (\ref{eq:kalappc5})
and marginalized over $(\gamma_A,\sigma_{AA})$ with $A \in \{ Q, S, \eta \}$.
Priors are uniform and run from 0.1 to 10 times the injected value per parameter.
The peak coincides with the injected values to within 
$\approx 0.080\,{\rm dex}$ and $\approx 0.034\,{\rm dex}$ 
for $\overline{Q}$ and $\overline{S}$ respectively.
The uncertainty (full width half maximum) amounts to
$\approx 0.021 \, {\rm dex}$ and $\approx 0.011 \, {\rm dex}$
for $\overline{Q}$ and $\overline{S}$ respectively.
Encouragingly, the error in the peak and the full width half maximum are small,
in the sense that both sit well inside the prior range
($\pm 1 \, {\rm dex}$).
The corner plot shows that the peak is biased marginally yet systematically away from the injected values,
in the sense that the error in the peak exceeds the full width half maximum
(exaggerated visually in the right panel of Figure \ref{fig:kalappc1}
by the fine horizontal scale).
Such a bias, although negligible in practice, is typical of 
certain nonlinear estimation algorithms \citep{gel74};
cf.\ the unbiased results of the linear Kalman filter in Figure \ref{fig:kal2}.
Its smallness engenders confidence in the approximations involved,
e.g.\ $\beta_3 = 0$.
Overall, the results in Figure \ref{fig:kalappc1} 
are comparable to Figures \ref{fig:kal1} and \ref{fig:kal2}.

\section{Refining the description of the disk-magnetosphere boundary
 \label{sec:kalappd}}
The accretion physics analyzed in this paper is presented deliberately in a simple form,
which adheres to the canonical magnetocentrifugal picture introduced
by \citet{gho79} and applied widely since \citep{pat21}.
There are three motivations for this approach.
(i) It is the first time in the literature that a Kalman filter is applied to estimate 
the parameters of accretion-powered pulsars, so it is prudent to present
the statistical technique simply for clarity.
(ii) Data sets available now and in the foreseeable future are limited to
$N \lesssim 10^3$ samples $P(t_n)$ and $L(t_n)$.
It is asking a lot of a Kalman filter to estimate more parameters
than the seven in \S\ref{sec:kal3a},
despite the encouraging test results in Figures \ref{fig:kal1}--\ref{fig:kal3}
and \S\ref{sec:kal4}.
(iii) There is no consensus in the literature about how to refine the canonical
magnetocentrifugal model through simple analytic modifications.
Impressive progress is occurring through three-dimensional
magnetohydrodynamic simulations
\citep{rom15},
but the output of such simulations does not take a form that feeds easily
into a Kalman filter or equivalent parameter estimation scheme.

That said, we emphasize that aspects of the
fluctuation dynamics in accretion-powered pulsars may not be represented
accurately by the white-noise, mean-reverting Langevin equations
(\ref{eq:kal5}) and (\ref{eq:kal9})--(\ref{eq:kal11}).
Data available today may not suffice to expose the inaccuracies,
as noted above,
but it is important to start investigating them in principle,
in anticipation of more data in the future.
In this appendix,
we take a preliminary first step towards refining the accretion model,
by adding deterministic descriptions of some of the complicated variability
at the disk-magnetosphere boundary,
which is absorbed in the Langevin dynamics in the main text.
We draw heavily on the theory of episodic accretion and trapped disks
\citep{dan10,dan12,dan17}
by way of illustration, and because it proposes convenient and
soundly motivated analytic scalings,
without seeking to privilege it over plausible and complementary
alternatives in the literature;
see \citet{lai14} for a review.

\subsection{Complexities at the boundary
 \label{sec:kalappda}}
The central approximation in
(\ref{eq:kal5}) and (\ref{eq:kal9})--(\ref{eq:kal11})
is that there exists a sharp boundary at radius $r=R_{\rm m}$
between a corotating dipole magnetosphere at $r < R_{\rm m}$
and a thin accretion disk at $r > R_{\rm m}$.
In reality,
the transition from magnetosphere to disk occurs gradually,
over an interaction region $R_{\rm m} - \Delta R \leq r \leq R_{\rm m}$,
with the disk surface density vanishing at $R_{\rm in}= R_{\rm m} - \Delta R$
\citep{dan10}.
This has implications for the hidden variables $Q(t)$, $S(t)$, and $\eta(t)$.
(i) In the interaction region,
the accretion flow is sheared and variable,
and the toroidal and vertical magnetic field components penetrating the region,
viz.\ $B_\phi$ and $B_z$ respectively,
where $(r,\phi,z)$ are cylindrical polar coordinates,
fluctuate in a complicated manner in space and time.
The scalar Maxwell stress $S(t)$ is inadequate by itself to capture the physics;
the ratio $\eta_B = | B_\phi/ B_z | \lesssim 1$
also plays an important role at the next level of approximation
\citep{wan87,spr93}.
(ii)
Vertical forces act to squeeze or inflate the disk,
to the point where it is not always thin.
For example,
the disk magnetic pressure $\sim B_\phi^2/(8\pi)$ can launch a vertical outflow,
opening up magnetic field lines
and disconnecting the disk at $r\gtrsim R_{\rm m}$ magnetically from the star
(before possibly reconnecting).
This modifies (\ref{eq:kal10}) for $S(t)$.
(iii) 
Radial forces in the interaction region decelerate the infall
and shut it off completely in the propeller regime $R_{\rm m } \gg R_{\rm c}$,
an effect which is absent from (\ref{eq:kal11}),
which maintains $\eta(t) \approx \eta_0$ for all $t$
irrespective of $R_{\rm m} / R_{\rm c}$.
The transition from free fall to propeller occurs abruptly 
over a transition length scale $\Delta R_2 \sim \Delta R$,
with $|R_{\rm m} - R_{\rm c}| \lesssim \Delta R_2$
and $\Delta R_2 \neq \Delta R$ in general
\citep{dan10}.
(iv)
In the regime $R_{\rm c} < R_{\rm m} \lesssim 1.3 R_{\rm c}$,
the disk-magnetosphere interaction may deposit enough angular momentum 
in the interaction region to pause the infall,
so that gas piles up near $R_{\rm m}$ without unbinding gravitationally
before it breaks through and accretes episodically onto the star,
viz.\ the trapped disk phenomenon
\citep{dan10,dan12,dan17}.
(v)
If enough energy and angular momntum are deposited in the interaction region,
parts of the disk at $r \gtrsim R_{\rm m}$ unbind,
launching a vertical outflow along twisted magnetic field lines
\citep{mat05,rom15}.
An outflow modifies (\ref{eq:kal9}) for $Q(t)$.
All the phenomena (i)--(v) are observed in three-dimensional magnetohydrodynamic simulations
\citep{rom15}.

In what follows, we investigate two rudimentary modifications of
(\ref{eq:kal5}) and (\ref{eq:kal9})--(\ref{eq:kal11}),
as a starting point for assessing the impact of the above phenomena
on the Kalman filter framework.
In Appendix \ref{sec:kalappdb},
we extend the torque and mass transfer laws (\ref{eq:kal5}) and (\ref{eq:kal11}) 
to include abrupt switching from the infall to the propeller regimes,
following the ``hyperbolic tangent'' approximation in \citet{dan17}.
Kalman filter results are generated for representative values of $\Delta R_2$.
In Appendix \ref{sec:kalappdc},
we analyze the trapped disk scenario in \citet{dan10} approximately
by modifying the definition (\ref{eq:kal3}) for $R_{\rm m}$
\citep{spr93},
adding a $\Delta R_2$-dependent transition to $\eta(t)$ via (\ref{eq:kal11}),
and adding a $\Delta R$-dependent term to the torque via (\ref{eq:kal5})
following \citet{dan12},
without solving self-consistently for the disk surface mass density
[cf.\ \citet{dan10,dan12}],
which is outside the scope of this paper and burdens the Kalman filter
with too many parameters to estimate.
Kalman filter results are generated for representative values of $\Delta R$ and $\Delta R_2$.
Finally, in Appendix \ref{sec:kalappdd},
we summarize qualitatively some of the issues raised by disk and stellar outflows
but do not model them mathematically,
as they lie outside the scope of the paper and the Kalman filter's capacity at present,
and analytic prescriptions like those in Appendices \ref{sec:kalappdb} and \ref{sec:kalappdc}
are hard to develop.

\subsection{Propeller transition
 \label{sec:kalappdb}}
As a starting point in modifying
(\ref{eq:kal5}) and (\ref{eq:kal9})--(\ref{eq:kal11}),
we incorporate approximately two aspects of the propeller transition.
First, we replace (\ref{eq:kal5}) with
\begin{equation}
 \frac{d\Omega}{dt}
 =
 I^{-1} (GM)^{1/2} R_{\rm m}(t)^{1/2} Q(t) \, {\rm tanh}
 \left[
 \frac{R_{\rm c}(t)-R_{\rm m}(t)}{\Delta R_2}
 \right]~,
\label{eq:kalappd1}
\end{equation}
which matches equation (3) in \citet{dan17}
up to a dimensionless proportionality constant of order unity,
which cannot be estimated independently by the Kalman filter
and is absorbed into $I^{-1} (GM)^{1/2}$.
The definitions (\ref{eq:kal3}) and (\ref{eq:kal4})
of $R_{\rm m}(t)$ and $R_{\rm c}(t)$ respectively remain unchanged.
Equation (\ref{eq:kalappd1}) describes the same sign change in the torque
as (\ref{eq:kal5}),
from $d\Omega / dt > 0$ for $R_{\rm m} < R_{\rm c}$
to $d\Omega / dt < 0$ for $R_{\rm m} > R_{\rm c}$,
but it allows the sign change to occur more abruptly,
if one has $\Delta R_2 \ll R_{\rm c}$.
The transition length scale $\Delta R_2$ is added to the parameter set ${\bf\Theta}$,
which the Kalman filter estimates.

Second, we replace the stochastic Langevin equation (\ref{eq:kal11})
with the deterministic algebraic equation
\begin{equation}
 \eta(t) = \frac{\eta_0}{2}
 \left\{
  1- {\rm tanh}
  \left[
   \frac{R_{\rm m}(t)-R_{\rm c}(t)}{\Delta R_2}
  \right]
 \right\}~.
\label{eq:kalappd2}
\end{equation}
Equation (\ref{eq:kalappd2}) ensures that the accretion onto the star
and its associated X-ray emission switch off in the strong propeller regime
$R_{\rm m} \gg R_{\rm c}$.
This is more realistic than (\ref{eq:kal11}),
where X-ray emission persists undiminished even for $R_{\rm m} \gg R_{\rm c}$,
with $| \eta(t) - \eta_0 | \lesssim \gamma_\eta^{-1/2} \sigma_{\eta\eta} \neq 0$.
Observations show that weakly magnetized low-mass X-ray binaries 
and strongly magnetized Be X-ray binaries have duty cycles
$\lesssim 3\%$ and $\lesssim 20\%$ respectively,
and $L(t)$ varies by $\lesssim 5$ orders of magnitude between outbursts and quiescence
\citep{klu14,yan15,dan17}.
In writing (\ref{eq:kalappd2}) instead of (\ref{eq:kal11}),
we assume implicitly that fluctuations in $\eta(t)$ and hence $L(t)$ are driven solely by
the propeller dynamics at the disk-magnetosphere boundary.
In reality, other radiative processes influence $\eta(t)$
independently of mass transfer.
We do not include them in (\ref{eq:kalappd2}) for two reasons.
One, the available X-ray data do not suffice to constrain more complicated models
with additional, radiative parameters.
Two, the radiative processes are not described by simple analytic expressions 
like (\ref{eq:kalappd2}),
which is why we resort to phenomenological mean reversion in (\ref{eq:kal11}).
Retaining (\ref{eq:kal11}) but replacing $\eta_0$ in (\ref{eq:kal11})
with the right-hand side of (\ref{eq:kalappd2}) would be problematic,
because white-noise fluctuations in (\ref{eq:kal11}) would send
$\eta(t)$ negative for $R_{\rm m} \gg R_{\rm c}$,
when the right-hand side of (\ref{eq:kalappd2}) vanishes.

\begin{figure}[ht]
\begin{center}
\includegraphics[width=11cm,angle=0]{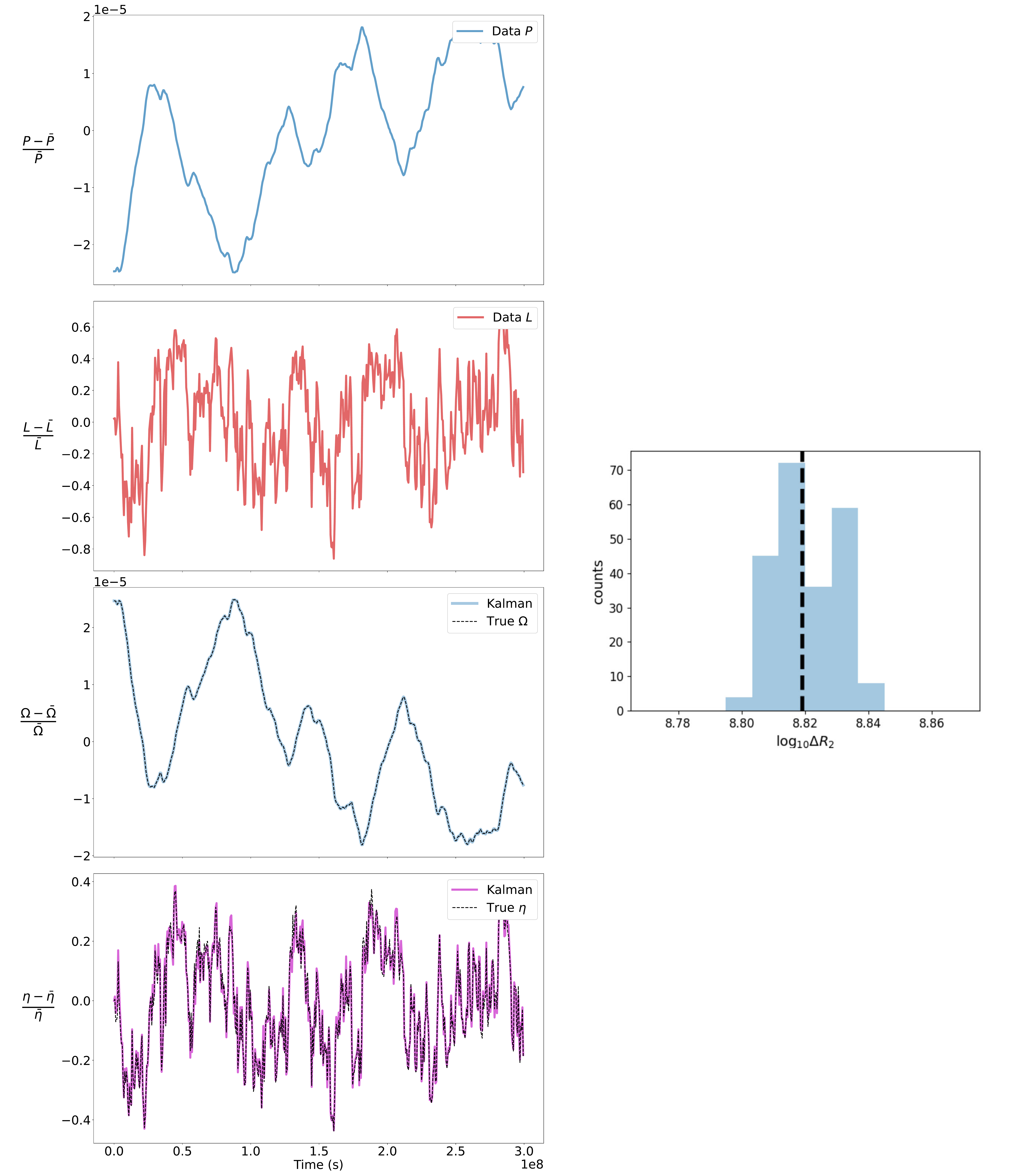}
\end{center}
\caption{
Kalman state tracking applied to a hypothetical accretion-powered pulsar 
with a propeller transition 
modeled by (\ref{eq:kalappd1}) and (\ref{eq:kalappd2})
with $\Delta R_2 = 0.09 R_{\rm m0}$.
Inputs:
synthetic measurements of spin period $P(t_n)$ (top panel at left)
and X-ray luminosity $L(t_n)$ (second panel at left)
versus time $t_n$ (units: ${\rm s}$), 
with $0< t_1 \leq \dots \leq t_{500} = 3.0\times 10^8\,{\rm s}$.
Outputs:
state variables
$\Omega(t_n)$ (third panel at left)
and
$\eta(t_n)$ (fourth panel at left)
versus time $t_n$ (units: ${\rm s}$);
the other state variables $Q(t_n)$ and $S(t_n)$
are omitted for clarity, as they resemble Figure \ref{fig:kal1}.
Parameter estimation:
posterior of $\Delta R_2$ (histogram; panel at right),
marginalized over the other six variables in ${\bf\Theta}$;
the injected value is indicated by the black, dashed, vertical line.
For the limited purpose of this test, 
$\gamma_A$ and $\sigma_{AA}$ with $A \in \{ Q,S \}$
are assumed known to accelerate the computation
but they would be estimated in general as in \S\ref{sec:kal4} and Appendix \ref{sec:kalappc}.
In the third and fourth panels at left,
colored, solid curves indicate the squared-error-minimizing,
reconstructed state sequence $\hat{\bf X}(t_n)$ generated by the Kalman filter,
and black, dashed curves indicate the true, injected state sequence ${\bf X}(t_n)$.
The vertical axes in the four panels at left display fractional and therefore
dimensionless quantities.
}
\label{fig:kalappd1}
\end{figure}

Figure \ref{fig:kalappd1} displays results from a repeat of the numerical experiment
in Figures \ref{fig:kal1} and \ref{fig:kal2} and Table \ref{tab:kal1},
except with (\ref{eq:kal5}) and (\ref{eq:kal11}) replaced by
(\ref{eq:kalappd1}) and (\ref{eq:kalappd2}) respectively,
and (\ref{eq:kal9}) and (\ref{eq:kal10}) replaced by
(\ref{eq:kalappc1}) and (\ref{eq:kalappc2}) respectively.
The revised model features seven parameters to be estimated, viz.\
${\bf\Theta} = (\beta_1,\beta_4,\beta_5,\gamma_Q,\gamma_S,\sigma_{QQ},\sigma_{SS})$,
with
\begin{equation}
 \beta_4
 =
 \frac{ (GM)^{1/3} }
  { \overline{\Omega}^{2/3} \Delta R_2 }~,
\label{eq:kalappd2a}
\end{equation}
\begin{equation}
 \beta_5
 =
 \frac{ (GM)^{1/5} \overline{Q}^{2/5} }
  { (4\pi)^{2/5} \overline{S}^{2/5} \Delta R_2 }~,
\label{eq:kalappd2b}
\end{equation}
$\beta_1$ defined according to (\ref{eq:kalappc4}),
and $\overline{\Omega}$ calculated directly from the data via the sample average (\ref{eq:kalappb1})
without assuming an equilibrium interpretation (see Appendix \ref{sec:kalappc}).
\footnote{
Once the seven components of ${\bf\Theta}$ are estimated,
we can solve (\ref{eq:kalappc4}), (\ref{eq:kalappd2a}), and (\ref{eq:kalappd2b})
for $\overline{Q}$, $\overline{S}$, and $\Delta R_2$,
given $\overline{\Omega}$ from the data.
We can then calculate $\overline{L}=\langle L(t) \rangle$ directly from the data 
via (\ref{eq:kalappb2})
and solve (\ref{eq:kalappc6}) for $\eta_0 = 2\overline{\eta}$.
Correlations between $Q(t)$ and $\eta(t)$ should be treated with caution.
In general, they introduce an extra parameter to be estimated,
as discussed in Appendix \ref{sec:kalappcb};
see (\ref{eq:kalappc7}).
}
We employ an unscented Kalman filter to analyze the nonlinear system,
as in Appendix \ref{sec:kalappc}
\citep{jul97},
with $N=5\times 10^2$.
We find that the Kalman filter tracks $\Omega(t)$ and $\eta(t)$ accurately;
compare the colored and dashed curves in the third and fourth panels on the left
of Figure \ref{fig:kalappd1}.
Likewise,
$Q(t)$ and $S(t)$ are tracked accurately;
they behave like in Figure \ref{fig:kal1}
(or equivalently Figure \ref{fig:kalappc1}) and are not displayed for brevity.
The system undergoes several propeller transitions,
e.g.\ from spin up to spin down at $t\approx 8 \times 10^7 \, {\rm s}$,
with $\Delta R_2 = 0.09 R_{\rm m0}$.
During the subsequent spin-down episode
$0.8 \lesssim t / (10^8 \, {\rm s}) \lesssim 1.2$,
we find $\eta(t) < \eta_0$ and hence $L(t) < \overline{L}=\langle L(t) \rangle$,
as marked by dips in the fourth and second panels respectively
on the left of Figure \ref{fig:kalappd1}.
Parameter estimation results are displayed in the right panel of Figure \ref{fig:kalappd1},
where we plot the posterior for $\Delta R_2$,
marginalized over the other six parameters
(which are taken as known artificially for the purpose of this test to accelerate the computation).
The error in the peak and the uncertainty (full width half maximum)
are given by $\approx 0.0067 \, {\rm dex}$ and $\approx 0.025 \, {\rm dex}$ respectively.
A preliminary analysis indicates that $N \geq 2\times 10^3$ samples are needed
to achieve convergence when scanning the complete parameter space,
above what is available typically from the current generation of X-ray timing experiments.
A full study of the estimation accuracy is postponed,
until larger volumes of astronomical data justify the revisions
(\ref{eq:kalappd1}) and (\ref{eq:kalappd2}).

In this section, we keep (\ref{eq:kalappc1}) to describe mean-reverting fluctuations in the
mass accretion rate driven by processes in the outer disk and companion star
\citep{lyu97},
noting that (\ref{eq:kalappc1}) does not incorporate outflows;
see Appendix \ref{sec:kalappdd}.
We also keep (\ref{eq:kalappc2}) for simplicity,
while noting that it is not self-consistent with (\ref{eq:kalappd2})
and should involve more than one tensor component in general.

\subsection{Disk trapping
 \label{sec:kalappdc}}
As a next step in modifying (\ref{eq:kal5}) and (\ref{eq:kal9})--(\ref{eq:kal11}),
we incorporate approximately the important phenomenon of disk trapping
\citep{dan10,dan12,dan17}.
In the weak propeller regime,
with $R_{\rm c} < R_{\rm m} \lesssim 1.3 R_{\rm c}$,
the complex disk-magnetosphere interaction deposits enough angular momentum in the inner disk
to inhibit accretion but not enough to unbind the disk material gravitationally
and drive a vertical outflow, cf. \citet{mat05}.
Instead, $R_{\rm m}(t)$ stalls and becomes almost independent of $Q(t)$,
so that the traditional magnetocentrifugal expression (\ref{eq:kal3})
for $R_{\rm m}(t)$ is modified
\citep{wan87,spr93}.
Gas piles up at the inner edge of the disk,
changing the disk density structure near $R_{\rm c}$ and hence the torque (\ref{eq:kal5}).
The accumulated gas eventually breaks through the magnetocentrifugal barrier,
e.g.\ via the Rayleigh-Taylor instability,
so that accretion onto the star occurs in episodic bursts,
which are shorter than accretion outbursts.
Hence under certain conditions the star spins down without an outflow,
even while some gas leaks onto the stellar surface,
loosening the torque-efficiency nexus implied by (\ref{eq:kalappd1}) and (\ref{eq:kalappd2}).

A comprehensive treatment of disk trapping involves calculating
the radial density profile of the disk near $R_{\rm c}$,
along the lines developed by \citet{dan10}.
Such a calculation does not fit easily into the mathematical framework of the Kalman filter
and introduces several new parameters,
which would challenge the estimation accuracy of the Kalman filter 
given current and prospective data volumes.
We therefore simplify the treatment using a subset of the analytic scalings
formulated by \citet{dan12}.
First, we replace the traditional formula (\ref{eq:kal3}) for $R_{\rm m}$
\citep{gho79}
with the following prescription for the inner disk radius,
$R_{\rm in} = R_{\rm m} - \Delta R$:
we solve 
\begin{equation}
 \frac{Q_{\rm co}(t)}{Q(t)}
 =
 \frac{1}{2}
 \left\{
  1- {\rm tanh}
  \left[
   \frac{R_{\rm in}(t)-R_{\rm c}(t)}{\Delta R_2}
  \right]
 \right\}~.
\label{eq:kalappd3}
\end{equation}
and
\begin{equation}
 R_{\rm in}(t)
 =
 \frac{\Omega(t) Q_{\rm co}(t)}{\pi S(t)}
\label{eq:kalappd4}
\end{equation}
simultaneously for $R_{\rm in}(t)$ and $Q_{\rm co}(t)$
as functions of $Q(t)$ and $S(t)$,
so that we can calculate the radiative efficiency
\begin{equation}
 \eta(t) 
 =
 \frac{Q_{\rm co}(t)}{Q(t)}
\label{eq:kalappd5}
\end{equation}
in terms of $Q(t)$ and $S(t)$.
Equations (\ref{eq:kalappd3}) and (\ref{eq:kalappd4}) resemble (\ref{eq:kalappd2});
see also equations (19) and (20) in \citet{dan10}.
They capture approximately how mass transfer onto the stellar surface ceases
in the propeller regime $R_{\rm in} > R_{\rm c}$.
However there are two subtleties of interpretation:
(i) $R_{\rm in} = R_{\rm m} - \Delta R$ replaces $R_{\rm m}$ in the right-hand side
of (\ref{eq:kalappd3});
and (ii) $Q_{\rm co}(t)$ is the mass accretion rate in the reference frame
comoving with $R_{\rm in}(t)$ at speed $\dot{R}_{\rm in}(t)$,
so we have $Q_{\rm co}(t) \neq Q(t)$ in general except at magnetocentrifugal equilibrium
($R_{\rm in} = R_{\rm c}$).
Equation (\ref{eq:kalappd4}) replaces (\ref{eq:kal3}).
Physically it describes the balance between the angular momentum flux
(equivalently the torque that makes the disk corotate with the star)
and the Maxwell stress at $R_{\rm in}$;
see equation (6) in \citet{dan10} and also \citet{spr93}.
Again there is a subtlety of interpretation:
here $S(t)$ is the $z\phi$-component of the Maxwell stress tensor,
which is $\eta_B = B_\phi/B_z$ times the $zz$-component in (\ref{eq:kal3}).
Assuming a dipole magnetic field and hence 
$S(t) = (4\pi)^{-1} \eta_B \mu^2 R_{\rm in}^{-6}$ inside the magnetosphere,
where $\mu$ is the star's magnetic moment,
we obtain from (\ref{eq:kalappd4}) the equivalent formula
\begin{equation}
 R_{\rm in}(t)
 =
 \left[
  \frac{\eta_B \mu^2}{4 \Omega(t) Q_{\rm co}(t)}
 \right]^{1/5},
\label{eq:kalappd6}
\end{equation}
which replaces (\ref{eq:kal8}) and agrees with equation (7) in \citet{dan10}.
Physically, the above description is consistent with the general observation by \citet{wan87},
that the growth rate of the toroidal field due to differential rotation
is proportional to $B_z$ [rather than $B_\phi$ as in \citet{gho79}],
which also ensures that the magnetic pressure $\sim B_\phi^2/(8 \pi)$
in the wound-up field remains lower than the thermal pressure
for $r \gtrsim R_{\rm c}$ and therefore does not disrupt the thin disk.

What is the torque law in the context of disk trapping?
As in the traditional picture
\citep{gho79},
the torque on the star comprises a spin-up component,
transmitted by the gas penetrating the magnetosphere and falling onto the stellar surface,
and a spin-down component,
which arises from the interaction between the magnetic field and the disk at $r\gtrsim R_{\rm c}$.
The two components are approximated by 
\citep{dan12,gen22}
\begin{equation}
 I \frac{d\Omega}{dt}
 =
 Q(t) [G M R_{\rm in}(t) ]^{1/2}
 -
 2\pi S(t) R_{\rm in}(t)^2 \Delta R 
 \left\{
  1 + {\rm tanh}
  \left[
   \frac{R_{\rm in}(t)-R_{\rm c}(t)}{\Delta R}
  \right]
 \right\}~,
\label{eq:kalappd7}
\end{equation}
which is the same as equation (12) in \citet{dan12}
and replaces (\ref{eq:kal5}) in the main text.
Physically (\ref{eq:kalappd7}) resembles (\ref{eq:kal5}),
in the sense that it implies $d\Omega / dt > 0$ for $R_{\rm in} < R_{\rm c}$,
where the hyperbolic tangent suppresses the magnetic torque,
and $d\Omega / dt < 0$ for $R_{\rm in} > R_{\rm c}$,
where the magnetic torque dominates.
However, there are some subtleties.
(i) The lever arm of the material component is $R_{\rm in}=R_{\rm m}-\Delta R$ 
rather than $R_{\rm m}$.
(ii) The width $\Delta R$ of the interaction region appears explicitly in the second term
on the right-hand side of (\ref{eq:kalappd7}) to allow for the scenario $\Delta R \ll R_{\rm c}$,
unlike in (\ref{eq:kal5}), 
where the factor $1-(R_{\rm m}/R_{\rm c})^{3/2}$ implicitly assumes $\Delta R \sim R_{\rm c}$.
(iii) The hyperbolic tangent smoothing function involves $\Delta R$ rather than $\Delta R_2$,
because the width of the interaction region is governed by related but different physics
to the propeller transition in Appendix \ref{sec:kalappdb}.
(iv) $S(t) \propto \eta_B \mu^2$ in (\ref{eq:kalappd7}) is the $z\phi$-component
of the Maxwell stress tensor, cf.\ the $zz$-component in (\ref{eq:kal5}).
(v) Zero torque occurs at $R_{\rm in} \approx R_{\rm c}$ 
but not $R_{\rm in} = R_{\rm c}$ exactly,
implying $\eta(t) \approx 0.5$ at zero torque.

\begin{figure}[ht]
\begin{center}
\includegraphics[width=11cm,angle=0]{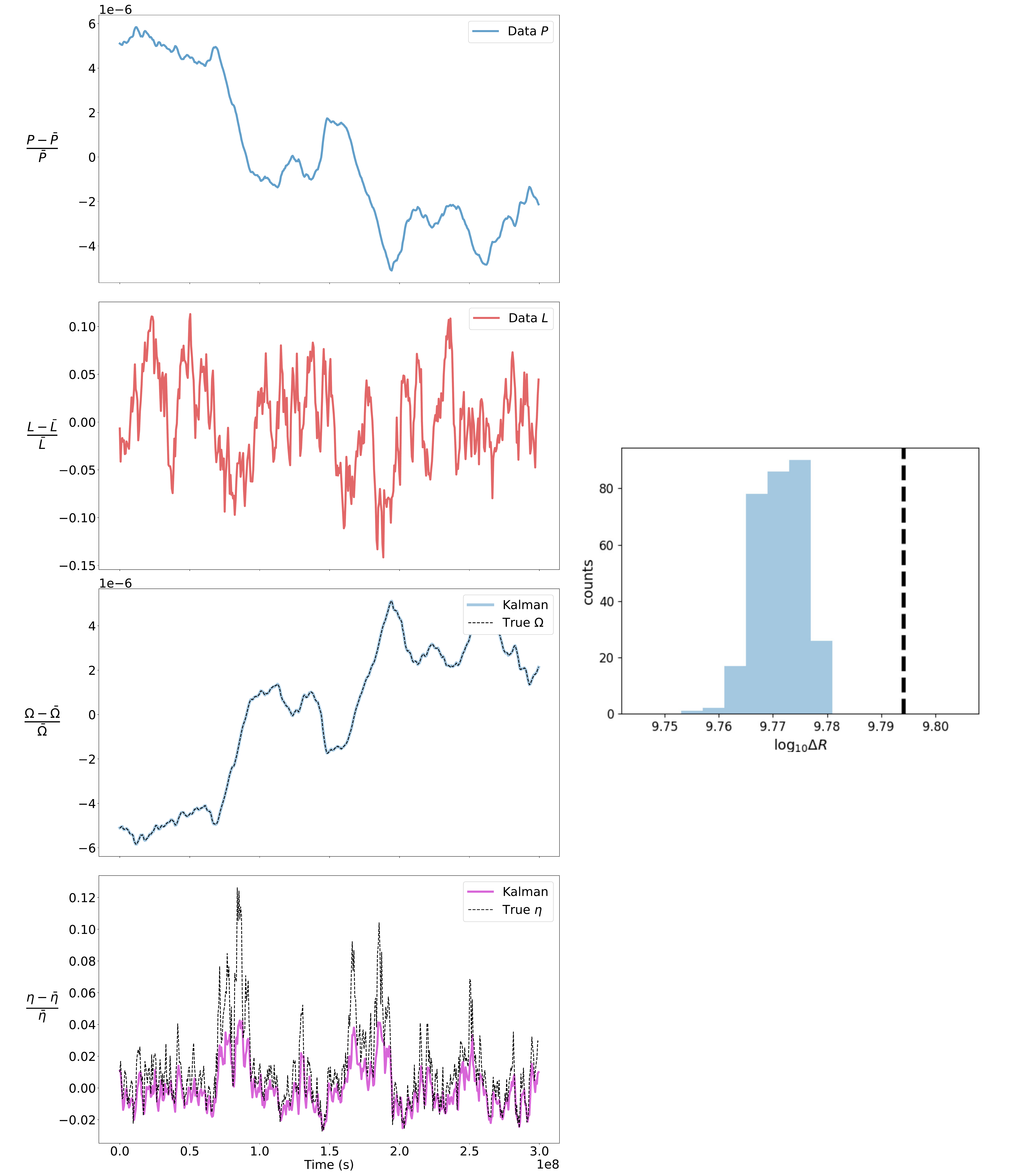}
\end{center}
\caption{
As for Figure \ref{fig:kalappd1}, but incorporating disk trapping modeled by
(\ref{eq:kalappd3})--(\ref{eq:kalappd7})
instead of (\ref{eq:kalappd1}) and (\ref{eq:kalappd2}),
with
$\Delta R = 0.80 R_{\rm m0}$ and $\Delta R_2 = 0.65 R_{\rm m0}$.
The axes of the four left-hand panels are the same as in Figure \ref{fig:kalappd1}.
The histogram in the right-hand panel displays
the marginalized posterior of $\Delta R$;
the injected value is indicated by the black, dashed, vertical line.
}
\label{fig:kalappd2}
\end{figure}

Figure \ref{fig:kalappd2} displays results from a repeat of the numerical experiment
in Figures \ref{fig:kal1}--\ref{fig:kal3} with disk trapping incorporated,
i.e.\ with (\ref{eq:kal5}) and (\ref{eq:kal11}) replaced by
(\ref{eq:kalappd7}) and (\ref{eq:kalappd5}) respectively,
supplemented by (\ref{eq:kalappd3}) and (\ref{eq:kalappd4}).
Equations (\ref{eq:kal9}) and (\ref{eq:kal10})
are also replaced by (\ref{eq:kalappc1}) and (\ref{eq:kalappc2})
to allow for disequilibrium.
The revised model features nine parameters ${\bf\Theta}$ to be estimated.
One possible combination is 
${\bf\Theta}=(\beta_6, \beta_7, \beta_8,\beta_9, \beta_{10},
  \gamma_Q, \gamma_S, \sigma_{QQ}, \sigma_{SS})$,
with
\begin{equation}
 \beta_6
 =
 \frac{(GM)^{1/2} \overline{Q}^{3/2}}
  {\pi^{1/2} I \overline{S}^{1/2} \overline{\Omega}^{1/2}}~,
\label{eq:kalappd10}
\end{equation}
\begin{equation}
 \beta_7
 =
 \frac{2 \overline{Q}^2 \overline{\Omega} \Delta R}
  {\pi I \overline{S}}~,
\label{eq:kalappd11}
\end{equation}
\begin{equation}
 \beta_8 
 =
 \frac{ \overline{\Omega} \, \overline{Q} }{ \pi \overline{S} \Delta R_2 }~,
\label{eq:kalappd12}
\end{equation}
\begin{equation}
 \beta_9
 =
 \frac{ (GM)^{1/3} }{ \overline{\Omega}^{2/3} \Delta R_2 }~,
\label{eq:kalappd13}
\end{equation}
and
\begin{equation}
 \beta_{10} 
 =
 \frac{\Delta R}{\Delta R_2}~.
\label{eq:kalappd14}
\end{equation}
We employ an unscented Kalman filter to do the nonlinear analysis
\citep{jul97},
as in Appendices \ref{sec:kalappc} and \ref{sec:kalappdb},
with $N=5\times 10^2$.
We find, as in Appendix \ref{sec:kalappdb}, that the Kalman filter tracks
$\Omega(t)$ and $\eta(t)$ accurately;
the solid, colored and dashed, black curves in the third and fourth
panels on the left of Figure \ref{fig:kalappd2} overlap closely.
The filter also tracks $Q(t)$ and $S(t)$ accurately;
the results resemble Figure \ref{fig:kalappc1} and are not plotted for brevity.
The graph of $\eta(t)$ (fourth panel on the left of Figure \ref{fig:kalappd2})
shows a smoother version of the episodic accretion seen in Figure 5 in \citet{dan12}.
``Dumping'' episodes occur during the intervals 
$0.7 \lesssim t/(10^8 \, {\rm s}) \lesssim 0.9$
and
$1.7 \lesssim t/(10^8 \, {\rm s}) \lesssim 1.9$,
accompanied by spikes in $\eta(t)$.
The dumping episodes are less pronounced than in \citet{dan12},
because the more realistic accretion model in \citet{dan12}
lets more gas pile up at the inner edge of the disk.
We deliberately choose $\Delta R=0.85 R_{\rm m0}$ and $\Delta R_2=0.65 R_{\rm m0}$ to be higher
than suggested by \citet{dan12}, in order to match the $\eta(t)$ behavior qualitatively.
\footnote{
Smaller values of $\Delta R$ and $\Delta R_2$ yield episodic accretion,
when one solves for the disk surface density profile self-consistently;
see {\S}3 and {\S}4 in \citet{dan10} and {\S}2 in \citet{dan12}.
However, it is challenging to incorporate the disk structure into the Kalman filter framework,
as discussed above, so we set $\Delta R$ and $\Delta R_2$ artificially high to compensate
for the purpose of testing.
\citet{dan12} argued that one has $\Delta R_2 \lesssim \Delta R < R_{\rm m0}$ typically,
and that $\Delta R$ must exceed a minimum threshold to suppress disruption
by the Kelvin-Helmholtz instability.
}
The estimation accuracy of the Kalman filter as a function of
$\gamma_Q$, $\gamma_S$, $\sigma_{QQ}$, $\sigma_{SS}$, and $\Delta R_2$
is studied in \S\ref{sec:kal4} and Appendix \ref{sec:kalappdb},
so here we focus on $\Delta R$.
The right panel of Figure \ref{fig:kalappd2} displays the marginalized posterior of $\Delta R$,
with the other parameters taken as known artificially to accelerate the computation.
We find that
the error in the peak and the uncertainty (full width half maximum)
are given by $\approx 0.020 \, {\rm dex}$ and $\approx 0.011 \, {\rm dex}$ respectively,
including a systematic but negligibly small bias 
like in Figure \ref{fig:kalappc1}.
A preliminary analysis indicates that $N\geq 2\times 10^3$ samples
are needed to achieve convergence when scanning the complete parameter space,
as in Appendix \ref{sec:kalappdb},
above what is available typically today.
A full study of the estimation accuracy is postponed, 
until larger volumes of astronomical data establish the need.

The results in Figure \ref{fig:kalappd2} are encouraging.
Nonetheless we emphasize that the revised equations of motion are still idealized.
Firstly,
(\ref{eq:kalappd5}) ascribes the efficiency $\eta(t)$ entirely to nonconservative mass transfer;
the radiative processes modeled phenomenologically through $\gamma_\eta$ and $\sigma_{\eta\eta}$
in (\ref{eq:kal11}) are missing.
Secondly,
we cannot track easily the disk density profile within the Kalman filter framework,
so we observe gentler episodic dumping of gas onto the stellar surface
than in \citet{dan10},
even though we capture the smoother disk trapping dynamics inherent 
in the torque and mass transfer laws (\ref{eq:kalappd3})--(\ref{eq:kalappd7}).
Thirdly,
in common with \citet{dan10},
we do not track the evolution of $B_\phi(t)$ and $B_z(t)$ separately,
nor do we resolve the system variables spatially in $\phi$ and $z$,
so we are blind to the rich phenomena observed in three-dimensional magnetohydrodynamic simulations,
such as finger-like accretion channels from Rayleigh-Taylor instabilities
\citep{rom03,rom05,zan13,rom15}.
Some of the latter physics may be captured crudely by the 
mean-reverting dynamics in (\ref{eq:kal9}) and (\ref{eq:kal10}),
parametrized by $\gamma_Q$, $\gamma_S$, $\sigma_{QQ}$, and $\sigma_{SS}$.
It remains to be seen, however,
whether white-noise fluctuations are representative of such processes
even as a rough approximation,
e.g.\ upon analyzing PSDs from the simulations;
see Figures 6 and 11 in \citet{rom21} for example.

\subsection{Outflows
 \label{sec:kalappdd}}
Three-dimensional magnetohydrodynamic simulations reveal that the inflow physics
in \S\ref{sec:kal2} and Appendices \ref{sec:kalappda}--\ref{sec:kalappdc}
is supplemented by outflows under a range of conditions;
see \citet{rom15} for a review.
Outflows can be launched by several mechanisms:
(i) magnetocentrifugal slingshot along open magnetic field lines anchored in the disk,
as long as they are inclined favorably
\citep{bla82};
(ii) episodic ejection of plasmoids,
when the magnetic pressure $\sim B_\phi^2/(8 \pi)$ inflates the disk
and severs temporarily its magnetic connection back to the star,
before reconnection restores it
\citep{lai14,uzd04};
(iii) magnetocentrifugal slingshot powered by the corotating magnetosphere
in the propeller regime;
(iv) field line bunching near $R_{\rm m}$,
which drives a conical, polar outflow for $R_{\rm m} < R_{\rm c}$
\citep{rom09,rom15};
and (v) outward redirection of an accretion flow along open magnetic field lines
anchored in the star rather than the disk
\citep{mat05,mat08,dan17}.

In order to incorporate an outflow into the Kalman filter framework, 
one needs an equation of motion for the mass ejection rate $Q_{\rm w}(t)$.
This is not easy to write down analytically;
$Q_{\rm w}(t)$ depends on the complicated magnetic topology and thermodynamics 
at the base of the outflow, which are not understood fully.
One may elect to bundle the complications into a mean-reverting Langevin equation,
analogous to (\ref{eq:kal9})--(\ref{eq:kal11}),
but this raises the number of parameters to be estimated from seven to nine
and may not be accurate physically.
Alternatively, one may draw upon phenomenological scalings in the literature,
e.g.\ equation (5) in \citet{mat05} for $Q_{\rm w}(t)$,
which introduces a new thermodynamic parameter.
Likewise, the spin-down torque exerted by a wind of type (v)
in the previous paragraph can be estimated phenomenologically as
\citep{mat08,dan17}
\begin{equation}
 \left.
 I\frac{d\Omega}{dt}
 \right|_{\rm w}
 \propto
 - Q_{\rm w}(t) \Omega(t) R_{\rm m}(t)^2~,
\label{eq:kalappd8}
\end{equation}
with
\begin{equation}
 R_{\rm m}(t)
 \propto
 [ Q_{\rm w}(t) / S(t) ]^{m'}
\label{eq:kalappd9}
\end{equation}
and $m' \approx 0.5$,
equivalent to equations (2) and (3) in \citet{mat08}
with $S \propto \mu^2 R_{\rm m}^{-6}$.
The above scalings raise the number of parameters from seven to 10.
Equations (\ref{eq:kalappd8}) and (\ref{eq:kalappd9}) have been generalized
semianalytically to interpolate between accretor ($R_{\rm m} < R_{\rm c}$)
and propeller ($R_{\rm m} > R_{\rm c}$) regimes
\citep{ire22}
and to model advection-dominated super-Eddington systems
\citep{cha19}
and plasmoid ejection
\citep{zan13}.

In light of the above challenges and in keeping with the scope of this paper,
we postpone the inclusion of outflows in the Kalman filter framework,
until larger data sets are available,
which justify estimating more parameters.

\section{Identifiability analysis
 \label{sec:kalappa}}
In an arbitrary Kalman filter,
the number of measurement variables 
(here two, namely $P$ and $L$)
does not necessarily equal the number of hidden state variables
(here four, namely $\Omega$, $Q$, $S$, and $\eta$,
for the linear model in \S\ref{sec:kal2e})
nor the number of system parameters
[here seven, namely
${\bf\Theta}=(\gamma_\Omega,\gamma_A,\sigma_{AA})$ with $A \in \{ Q, S, \eta \}$,
for the linear model in \S\ref{sec:kal2e}].
Several scenarios are therefore possible.
Sometimes the system parameters can be inferred uniquely,
even when there are fewer measurement variables than hidden state variables.
On other occasions the opposite holds:
some parameters cannot be inferred uniquely,
no matter how plentiful the data are,
because they enter through combinations that cannot be disentangled,
even when there are more measurement variables than hidden state variables.
What scenario applies to any specific problem can be determined
by performing a formal {\em identifiability analysis}
on the dynamical and measurement equations of the Kalman filter.
An identifiability analysis is a standard tool in
electrical engineering
\citep{bel70}.
In this appendix we apply it to the linearized system in \S\ref{sec:kal2e}
and find that all seven system parameters
${\bf\Theta}=(\gamma_{\Omega}, \gamma_A,\sigma_{AA})$ with $A \in \{ Q, S, \eta \}$
can be identified from the measured time series $P(t_n)$ and $L(t_n)$.
Identifiability analyses for the generalized, nonlinear models 
in Appendices \ref{sec:kalappc} and \ref{sec:kalappd} are postponed,
until data volumes grow to the point where such models are applied in practice.

To test for identifiability,
we must count the number of independent constraints imposed on ${\bf\Theta}$
by the data, acting through the Kalman recursion relations in \S\ref{sec:kal3b}.
We begin the analysis in the absence of noise.
Transforming temporarily from discrete to continuous time for the sake
of notational convenience,
we have 
$\dot{\bf X} = {\bf AX}$ and ${\bf Y} = {\bf CX}$,
with the $4\times 4$ matrix ${\bf A}$ and $2\times 4$ matrix ${\bf C}$
defined in \S\ref{sec:kal3b}.
We now seek to write ${\bf X}$ in terms of ${\bf Y}$.
As the rank of ${\bf Y}$ is less than the rank of ${\bf X}$,
we must supplement ${\bf Y}$ with its derivatives $\dot{\bf Y}$ and $\ddot{\bf Y}$.
In discrete time, this is equivalent to supplementing $P(t_n)$ and $L(t_n)$
with $P(t_{n-1})$, $L(t_{n-1})$, $P(t_{n-2})$, and $L(t_{n-2})$.
In continuous time, we obtain
$\dot{\bf Y} = {\bf C} \dot{\bf X} = {\bf CAX}$
and
$\ddot{\bf Y} = {\bf CA}^2{\bf X}$.
Upon combining the expressions for ${\bf Y}$ and $\dot{\bf Y}$
in terms of ${\bf X}$,
we obtain four independent, linear equations, viz.\
\begin{equation}
 \left(
  \begin{tabular}{c}
   $P_1$ \\ $L_1$ \\ $\dot{P}_1$ \\ $\dot{L}_1$
  \end{tabular}
 \right)
 =
 \left(
  \begin{tabular}{cccc}
   $-1$ & 0 & 0 & 0 \\ 
   0 & 1 & 0 & 1 \\
   $\gamma_\Omega$ & $\lambda_0$ & 
     $-\lambda_0$ & 0 \\
  0 & $-\gamma_Q$ & 0 & $-\gamma_\eta$
  \end{tabular}
 \right)
 \left(
  \begin{tabular}{c}
   $\Omega_1$ \\ $Q_1$ \\ $S_1$ \\ $\eta_1$
  \end{tabular}
 \right)~,
\label{eq:kalappa1}
\end{equation}
with $\lambda_0 = 3\gamma_\Omega / 5$.
The $4\times 4$ matrix on the right-hand side of (\ref{eq:kalappa1})
is invertible for $\gamma_Q \neq \gamma_\eta$,
whereupon we can solve for $\Omega_1$, $Q_1$, $S_1$, and $\eta_1$ 
in terms of $P_1$, $L_1$, $\dot{P}_1$, and $\dot{L}_1$ or,
in discrete terms,
$P_1(t_n)$, $L_1(t_n)$, $P_1(t_{n-1})$, and $L_1(t_{n-1})$.
That is, all four hidden state variables can be recovered as functions of time
from the measured time series.

To check which of the system parameters ${\bf \Theta}$ are identifiable,
we evaluate $\ddot{\bf Y} = {\bf CA}^2{\bf X}$. The matrix algebra
is straightforward and yields
\begin{equation}
 \left(
  \begin{tabular}{c}
   $\ddot{P}_1$ \\ $\ddot{L}_1$
  \end{tabular}
 \right)
 =
 \left(
  \begin{tabular}{cccc}
   $-\gamma_\Omega^2$ & 
    $-\lambda_0(\gamma_\Omega+\gamma_Q)$ & 
    $\lambda_0(\gamma_\Omega+\gamma_S)$ & 0 \\ 
   0 & $\gamma_Q^2$ & 0 & $\gamma_\eta^2$
  \end{tabular}
 \right)
 \left(
  \begin{tabular}{c}
   $\Omega_1$ \\ $Q_1$ \\ $S_1$ \\ $\eta_1$
  \end{tabular}
 \right)~.
\label{eq:kalappa2}
\end{equation}
Rewriting the hidden state variables in terms of
$P_1$, $L_1$, $\dot{P}_1$, and $\dot{L}_1$ with the aid of (\ref{eq:kalappa1}),
we arrive at
\begin{equation}
 \left(
  \begin{tabular}{c}
   $\ddot{P}_1$ \\ $\ddot{L}_1$
  \end{tabular}
 \right)
 =
 \left(
  \begin{tabular}{cccc}
   $-\gamma_S \gamma_\Omega$ & 
    $\frac{\lambda_0 \gamma_\eta (\gamma_Q-\gamma_S)}{\gamma_Q - \gamma_\eta}$ & 
    $-(\gamma_\Omega+\gamma_S)$ &
    $\frac{\lambda_0 (\gamma_Q-\gamma_S)}{\gamma_Q - \gamma_\eta}$ \\
   0 & $-\gamma_\eta \gamma_Q$ & 0 & $-(\gamma_Q+\gamma_\eta)$
  \end{tabular}
 \right)
 \left(
  \begin{tabular}{c}
   $P_1$ \\ $L_1$ \\ $\dot{P}_1$ \\ $\dot{L}_1$
  \end{tabular}
 \right)~.
\label{eq:kalappa3}
\end{equation}
Equation (\ref{eq:kalappa3}) is a system of ordinary differential equations
involving measurement variables and their derivatives only,
namely $P_1$, $L_1$, $\dot{P}_1$, $\dot{L}_1$, $\ddot{P}_1$, and $\ddot{L}_1$.
Hence all six nonzero elements of the $2\times 4$ matrix on the right-hand side 
can be estimated (``identified'') with enough data.
It is easy to resolve the six measured elements into
$\gamma_Q$, $\gamma_S$, and $\gamma_\eta$,
along with $\gamma_\Omega$ (which involves equilibrium quantities only).

The above analysis is performed on the noise-free Kalman equations.
It does not guarantee that the noise amplitudes
$\sigma_{QQ}$, $\sigma_{SS}$, and $\sigma_{\eta\eta}$ are identifiable.
Experience across many electrical engineering applications
suggests that noise amplitudes are usually identifiable
\citep{bel70},
because they are transformed versions of the dispersions
of the measurement variables, 
e.g.\ $\langle [ P(t_n) - \langle P(t_n) \rangle ]^2 \rangle^{1/2}$.
To check this formally for the system in \S\ref{sec:kal2e},
we calculate the covariance matrices ${\bf V}$ and ${\bf W}$
of the hidden state and measurement variables respectively,
with components
$V_{ij} = \langle \delta X_i(t) \delta X_j(t) \rangle$,
$W_{ij} = \langle \delta Y_i(t) \delta Y_j(t) \rangle$,
$\delta X_i(t) = X_i(t) - \langle X_i(t) \rangle$,
and
$\delta Y_i(t) = Y_i(t) - \langle Y_i(t) \rangle$,
assuming nonzero process noise but zero measurement noise.
We then count the number of independent parameters in ${\bf V}$ and compare
with the number of independent pieces of measured information in ${\bf W}$.

The covariance matrices are defined as usual by
\citep{gar94}
\begin{equation}
 {\bf V}
 =
 \int_0^t dt' \, \exp({\bf A}t') {\bf \Sigma} \exp({\bf A}^{\rm T} t')
\label{eq:kalappa4}
\end{equation}
and
\begin{equation}
 {\bf W} = {\bf C} {\bf V} {\bf C}^{\rm T}~,
\label{eq:kalappa5}
\end{equation}
with 
${\bf \Sigma} = 
 {\rm diag}(0,\sigma_{QQ}^2/Q_0^2 , \sigma_{SS}^2 / S_0^2 , \sigma_{\eta\eta}^2 / \eta_0^2)$
for the special case of uncorrelated white noise in (\ref{eq:kal12}).
The superscript T denotes the matrix transpose.
Recall that $Q_0$, $S_0$, and $\eta_0$ are expressible in terms of $\gamma_\Omega$
through (\ref{eq:kalappb4})--(\ref{eq:kalappb6}).
It is straightforward to evaluate (\ref{eq:kalappa4}) and (\ref{eq:kalappa5})
for the $4\times 4$ matrix ${\bf A}$ and $2\times 4$ matrix ${\bf C}$
defined in \S\ref{sec:kal3b}. 
We write down the nonzero, independent components for reference as follows:
\begin{eqnarray}
 V_{\Omega\Omega}
 & = &
 \frac{\lambda_0^2 \sigma_{QQ}^2}{2 Q_0^2}
 \left[
  -\frac{(-1+e^{-2\gamma_\Omega t})}{\gamma_\Omega (\gamma_\Omega - \gamma_Q)^2}
  -\frac{(-1+e^{-2\gamma_Q t})}{\gamma_Q (\gamma_\Omega - \gamma_Q)^2}
  -\frac{4(1-e^{-\gamma_\Omega t-\gamma_Q t})}{(\gamma_\Omega + \gamma_Q) (\gamma_\Omega - \gamma_Q)^2}
 \right]
 \nonumber \\
 & & 
 + \frac{\lambda_0^2 \sigma_{SS}^2}{2 S_0^2}
 \left[
  -\frac{(-1+e^{-2\gamma_\Omega t})}{\gamma_\Omega (\gamma_\Omega - \gamma_S)^2}
  -\frac{(-1+e^{-2\gamma_S t})}{\gamma_S (\gamma_\Omega - \gamma_S)^2}
  -\frac{4(1-e^{-\gamma_\Omega t-\gamma_S t})}{(\gamma_\Omega + \gamma_S) (\gamma_\Omega - \gamma_S)^2}
 \right]~,
\label{eq:kalappa6}
 \\
 V_{\Omega Q}
 & = &
 \frac{\lambda_0 \sigma_{QQ}^2}{2 (\gamma_\Omega-\gamma_Q) Q_0^2}
 \left[
  \frac{-1+e^{-2\gamma_Q t}}{\gamma_Q}
  + \frac{2(1-e^{-\gamma_\Omega t-\gamma_Q t})}{\gamma_\Omega + \gamma_Q}
 \right]~,
\label{eq:kalappa7}
 \\
 V_{\Omega S}
 & = &
 \frac{\lambda_0 \sigma_{SS}^2}{2 (\gamma_\Omega-\gamma_S) S_0^2}
 \left[
  \frac{1-e^{-2\gamma_S t}}{\gamma_S}
  + \frac{2(-1+e^{-\gamma_\Omega t-\gamma_S t})}{\gamma_\Omega + \gamma_S}
 \right]~,
\label{eq:kalappa8}
 \\
 V_{QQ}
 & = &
 \frac{(1-e^{-2 \gamma_Q t}) \sigma_{QQ}^2}{2 \gamma_Q Q_0^2}~,
\label{eq:kalappa9}
 \\
 V_{SS}
 & = &
 \frac{(1-e^{-2 \gamma_S t}) \sigma_{SS}^2}{2\gamma_S S_0^2}~,
\label{eq:kalappa10}
 \\
 V_{\eta\eta}
 & = &
 \frac{(1-e^{-2 \gamma_\eta t}) \sigma_{\eta\eta}^2}{2\gamma_\eta \eta_0^2}~,
\label{eq:kalappa11}
\end{eqnarray}
and
\begin{eqnarray}
 W_{PP}
 & = &
 V_{\Omega\Omega}~,
\label{eq:kalappa12}
 \\
 W_{PL}
 & = &
 - V_{\Omega Q}~,
\label{eq:kalappa13}
 \\
 W_{LL}
 & = &
 V_{QQ} + V_{\eta\eta}~.
\label{eq:kalappa14}
\end{eqnarray}
Upon combining (\ref{eq:kalappa6})--(\ref{eq:kalappa14}),
we arrive at the set of linear equations
\begin{equation}
 \left(
  \begin{tabular}{c}
   $W_{PP}$ \\ $W_{PL}$ \\ $W_{LL}$
  \end{tabular}
 \right)
 =
 \left(
  \begin{tabular}{ccc}
   $M_{11}$ & $M_{12}$ & 0 \\
   $M_{21}$ & 0 & 0 \\
   $M_{31}$ & 0 & $M_{33}$
  \end{tabular}
 \right)
 \left(
  \begin{tabular}{c}
   $\sigma_{QQ}^2 / Q_0^2$ \\ $\sigma_{SS}^2 / S_0^2$ \\ $\sigma_{\eta\eta}^2 / \eta_0^2$
  \end{tabular}
 \right)~,
\label{eq:kalappa15}
\end{equation}
with
\begin{eqnarray}
 M_{11}
 & = & 
 \frac{\lambda_0^2}{2}
 \left[
  -\frac{-1+e^{-2\gamma_\Omega t}}{\gamma_\Omega (\gamma_\Omega - \gamma_Q)^2}
  -\frac{-1+e^{-2\gamma_Q t}}{\gamma_Q (\gamma_\Omega - \gamma_Q)^2}
  -\frac{4(1-e^{-\gamma_\Omega t-\gamma_Q t})}{(\gamma_\Omega + \gamma_Q) (\gamma_\Omega - \gamma_Q)^2}
 \right]~,
\label{eq:kalappa16}
 \\
 M_{12}
 & = & 
 \frac{\lambda_0^2}{2}
 \left[
  -\frac{-1+e^{-2\gamma_\Omega t}}{\gamma_\Omega (\gamma_\Omega - \gamma_S)^2}
  -\frac{-1+e^{-2\gamma_S t}}{\gamma_S (\gamma_\Omega - \gamma_S)^2}
  -\frac{4(1-e^{-\gamma_\Omega t-\gamma_S t})}{(\gamma_\Omega + \gamma_S) (\gamma_\Omega - \gamma_S)^2}
 \right]~,
\label{eq:kalappa17}
 \\
 M_{21}
 & = &
 - \frac{\lambda_0}{2(\gamma_\Omega-\gamma_Q)}
 \left[
  \frac{-1+e^{-2\gamma_Q t}}{\gamma_Q}
  + \frac{2(1-e^{-\gamma_\Omega t-\gamma_Q t})}{\gamma_\Omega + \gamma_Q}
 \right]~,
\label{eq:kalappa18}
 \\
 M_{31} 
 & = &
 \frac{1 - e^{-2\gamma_Q t}}{2\gamma_Q }~,
\label{eq:kalappa19}
 \\
 M_{33} 
 & = &
 \frac{1 - e^{-2\gamma_\eta t}}{2\gamma_\eta}~.
\label{eq:kalappa20}
\end{eqnarray}
The $3\times3$ matrix in (\ref{eq:kalappa15}) is invertible,
so the noise amplitudes can be solved in terms of the data in ${\bf W}$,
together with $\gamma_Q$, $\gamma_S$, $\gamma_\eta$, and $\gamma_\Omega$ from (\ref{eq:kalappa3}),
and the equilibrium state.
That is, $\sigma_{QQ}$, $\sigma_{SS}$, and $\sigma_{\eta\eta}$ are identifiable in principle.

Ultimately identifiability must be verified empirically for a finite data set.
The results in \S\ref{sec:kal4} imply that
the parameters ${\bf \Theta}= (\gamma_\Omega,\gamma_A,\sigma_{AA})$
with $A\in \{ Q, S, \eta \}$ are identifiable in practice
for typical data volumes in the application studied in this paper.

As an aside of physical interest, equations (\ref{eq:kalappa6})--(\ref{eq:kalappa11})
imply that the spin fluctuations $\Omega_1$ are much smaller in magnitude than the
fluctuations $Q_1$, $S_1$, and $\eta_1$ of the hidden state variables,
as seen in Figure \ref{fig:kal1} and discussed in \S\ref{sec:kal4b}.
A typical accretion-powered pulsar has $\gamma_\Omega \ll \gamma_A$
with $A \in \{ Q, S, \eta \}$.
Physically, torque fluctuations of a given fractional amplitude drive
spin fluctuations of a smaller fractional amplitude,
because the star's moment of inertia is large.
Evaluating (\ref{eq:kalappa6})--(\ref{eq:kalappa11}) in the regime $\gamma_\Omega \ll \gamma_A$,
we obtain the leading-order scalings
$V_{\Omega\Omega} \sim (\gamma_\Omega / \gamma_Q^2) (\sigma_{QQ}^2 / Q_0^2) +
 (\gamma_\Omega / \gamma_S^2) (\sigma_{SS}^2 / S_0^2)$,
$V_{QQ} \sim \sigma_{QQ}^2 / (\gamma_Q Q_0^2)$,
$V_{SS} \sim \sigma_{SS}^2 / (\gamma_S S_0^2)$,
and
$V_{\eta\eta} \sim \sigma_{\eta\eta}^2 / (\gamma_\eta \eta_0^2)$.
The scalings imply 
$V_{\Omega\Omega} / V_{QQ} \sim 
 \max [ \gamma_\Omega / \gamma_Q, 
 (\gamma_\Omega / \gamma_S) (\gamma_Q / \gamma_S) (\sigma_{SS}^2/S_0^2) (\sigma_{QQ}^2 / Q_0^2)^{-1} ] \ll 1$,
consistent with Figure \ref{fig:kal1};
the variances $V_{\Omega\Omega}$ and $V_{QQ}$ measure the characteristic magnitudes
of the fluctuations $\Omega_1$ and $Q_1$ respectively.
A similar conclusion follows
for $V_{\Omega\Omega} / V_{SS}$ and $V_{\Omega\Omega} / V_{\eta\eta}$.

\end{document}